\documentclass[aps,prb,amsmath,amssymb,amsfonts,superscriptaddress,floatfix]{revtex4} 
\pdfoutput=1
\usepackage{amsmath,amsthm,amsfonts,amssymb}
\usepackage{graphicx}

\usepackage{color}
\usepackage{amsthm}
\numberwithin{equation}{section}
\numberwithin{figure}{section}

\newtheorem{theorem}{Theorem}
\newtheorem{lemma}{Lemma}
\newtheorem{corollary}{Corollary}

\newtheorem{definition}{Definition}
\newtheorem*{unth}{Theorem}

\newcommand{\be}{\begin{equation}}
\newcommand{\ee}{\end{equation}}

\newcommand{\wt}{{\rm wt}}

\begin{document}

\title{Weight Reduction for Quantum Codes}
\begin{abstract}

\end{abstract}
\author{Matthew B.~Hastings}

\affiliation{Station Q, Microsoft Research, Santa Barbara, CA 93106-6105, USA}
\affiliation{Quantum Architectures and Computation Group, Microsoft Research, Redmond, WA 98052, USA}

\begin{abstract}
We present an algorithm that takes a CSS stabilizer code as input, and outputs another CSS stabilizer code such that the stabilizer generators all have weights $O(1)$ and such that $O(1)$ generators
act on any given qubit.  The number of logical qubits is unchanged by the procedure, while we give bounds on the increase in number of physical qubits and in the effect on distance and other code parameters, such as soundness (as a locally testable code) and ``cosoundness" (defined later).  Applications are discussed, including to codes from high-dimensional manifolds which have logarithmic weight stabilizers.  Assuming a conjecture in geometry\cite{hdm}, this allows the construction of CSS stabilizer codes with generator weight $O(1)$ and almost linear distance.  Another application of the construction is to increasing the distance to $X$ or $Z$ errors, whichever is smaller, so that the two distances are equal.
\end{abstract}
\maketitle

\section{Introduction}
Many different constructions of CSS\cite{css} quantum stabilizer codes have been given.  However, a major problem has been obtaining simultaneously a large distance with low weight stabilizer generators,
where the weight of a stabilizer generator is the number of qubits on which it acts.
If we require that stabilizer generators have
weight which is $O(1)$, then 
the toric code achieves distance $\Theta(\sqrt{N})$, where $N$ is the number of physical qubits.   The best known distance improves on this only by a polylogarithm: the best known
known distance for a quantum code family is $\Theta(\sqrt{N \log(N)})$ as in Ref.~\onlinecite{fml}.
This contrasts with the classical case where it is possible to obtain linear distance with $O(1)$ weight stabilizer generators\cite{ldpc}.

On the other hand, if we allow the stabilizer generator weight to increase, then it is also possible to improve the distance.
Using a random code, one can obtain distance $\Theta(N)$ with stabilizer generator weight $\Theta(N)$, while using a ``homological product"\cite{fh,bh,hp} of random codes, it is possible
to obtain distance $\Theta(N)$ with stabilizer generator weight  $\Theta(\sqrt{N})$ as in Ref.~\onlinecite{bh}, and it is conjectured that an extension of that procedure would reduce the stabilizer
generator weight to $\Theta(N^{\alpha})$ for some values of $\alpha$ which is smaller than $1/2$ (the strongest conjecture would allow arbitrarily small $\alpha$ however proving such a conjecture is likely
to be difficult).
An alternative construction,  using a toric code on high dimensional random tori\cite{hdm} gives stabilizer generator weight $\Theta(\log(N))$ with distance $\Theta(N^{1-\epsilon})$ for any $\epsilon>0$, assuming 
a conjecture in geometry.  Finally, it is possible to obtain distance $\Theta(\sqrt{N})$ with stabilizer weight $O(1)$ and linear number of encoded bits\cite{tz}.

Thus, a natural question is whether one can reduce this stabilizer generator weight $N^{\alpha}$ or $\log(N)$ to $O(1)$, without hurting the distance properties of the code too strongly.
In this paper, we give a weight reduction procedure that does.  This procedure takes as input a CSS quantum code whose stabilizer generators may have high weight (i.e., they may act on a large number of qubits) and whose qubits may participate in many stabilizer generators.  The result is a new CSS quantum code where all stabilizer generators have
weight $O(1)$ and all qubits participate in at most $O(1)$ stabilizer generators.  This procedure does not change the number of logical qubits, but can unfortunately increase the number
of physical qubits $N$ required and reduce the distance of the code.  However, the effects on $N$ and distance can be controlled based on the parameters of the original code.

This procedure is distinct from the weight reduction procedure in Ref.~\onlinecite{subsyslin}, as the procedure of Ref.~\onlinecite{subsyslin} yields a subsystem code while the procedure here yields a stabilizer code.  Unfortunately, the results of the procedure will not be quite as strong as a result: while Ref.~\onlinecite{subsyslin} yields a subsystem code with distance $\Theta(N^{1-\epsilon})$ when applied to a concatenated code, the results here will not give a good distance when applied to a concatenated code.

Before giving the main theorem and applications of this theorem, we review some notation.
Throughout this paper, we consider the case of qubit codes.  The results generalize directly to qudits with minor changes.  We comment in some places on these changes; if not explicitly mentioned, we assume the qubit case.
Throughout this paper, when we refer to a ``generator", this refers to a generator of the stabilizer group.  A product of such generators will be called an element of the stabilizer group.
Define a $Z$-type operator to be a product of Pauli $Z$ operators and define
an $X$-type operator to be a product of Pauli $X$ operators.
We consider a CSS code, so that the generators are of two types, $X$-type and $Z$-type.  
We let $N$ denote the number of physical qubits and $K$ denote the number of logical qubits.
The weight of an $X$-type or $Z$-type operator $O$, written $\wt(O)$,
is equal to the number of Pauli operators that appear in the product defining that operator $O$.
We define several parameters to quantify the generators.
Let $w_X$ denote the maximum weight of $X$-type generator and let $w_Z$ denote the maximum weight of $Z$-type generator.
Let $q_X$ denote the maximum over all qubits of the number of $X$-type generators acting on that qubit and let
$q_Z$ denote the maximum over all qubits of the number of $Z$-type generators acting on that qubit.
Let $n_X$ and $n_Z$ denote the number of $X$-type and $Z$-type generators, respectively.
Let $d_X$ denote the minimum weight of an $X$-type logical operator (i.e., an $X$-type operator that is not in the stabilizer group but that commutes with all $Z$-type generators), and $d_Z$ denote the minimum weight of a $Z$-type logical operator.
Finally, we use $W_X$ and $W_Z$ to denote the sum of weights of $X$-type and $Z$-type generators, respectively; we have the trivial bound that $W_X\leq n_X w_X$ and $W_Z \leq n_Z w_Z$.

Code families with $w_X,w_Z$ both $O(1)$ are called quantum LDPC codes.  We further define
\begin{definition}
A code family with $w_X,w_Z,q_X,q_Z$ all $O(1)$ is called ``strongly LDPC".
\end{definition}

Finally, we define the soundness parameter of a quantum code considered as a quantum locally testable code\cite{qltc,tzqltc}:
\begin{definition}
\label{soundness}
Let $w$ be an integer.  Define $\epsilon_Z(w)$ as follows.
Consider the infimum over all $Z$-type operators $O$, such that $O$ has weight $w$ and such that $O$ does not commute with at
least one stabilizer, of the following quantity: take the maximum, over all $Z$-type operators $P$ which commute with all stabilizers, of the ratio of the number of generators which do not commute with $O$ to the weight of $OP$.  This infimum is $\epsilon_Z(w)$.
Define $\epsilon_Z={\rm min}_{w>0} \epsilon_Z(w)$.

Define $\epsilon_X(w)$ and $\epsilon_X$ similarly, with $X,Z$ interchanged everywhere.
\end{definition}

\subsection{Main Theorem and Applications}
We now give the main theorem.  While this theorem involves some complicated-seeming equations to determine the asymptotic parameters of the new code in terms of the
asymptotic parameters of the original code, the details of these parameters are not too important.  Below we discuss some applications where the calculation of the parameters simplifies.
Indeed, in the application to codes from high dimensional manifolds with $w_X=w_Z=q_X=q_Z=O(\log(N))$, the distance $d_X,d_Z$ are reduced only by a polylogarithmic factor and in the proof of the intermediate lemmas many of the estimates simplify.
For most applications of this theorem, one will want to take $\epsilon$ small; if instead one takes the limit $\epsilon\rightarrow +\infty$, one finds $\tau_X+\tau_Z \rightarrow 1$.
\begin{theorem}
\label{mainth}
Let $\epsilon$ be any positive constant.
Consider a family of quantum codes, with
$w_X=O(N^{\alpha_X}),w_Z=O(N^{\alpha_Z}),q_X=O(N^{\beta_X}),q_Z=O(N^{\beta_Z}),W_X=O(N^{\sigma_X}),W_Z=O(N^{\sigma_Z}),d_X=\Omega(N^{\tau_X}),d_Z=\Omega(N^{\tau_Z})$.
Assume that $\sigma_X,\sigma_Z\geq 1$.
Then, there exists a quantum code family $C^{new}$ which is strongly LDPC with 
$d^{new}_X=\Omega((N^{new})^{\tau^{new}_X}),d^{new}_Z=\Omega((N^{new})^{\tau^{new}_Z})$, with
\be
\tau^{new}_X=\frac{\tau_X+\epsilon\alpha_X+(1+\epsilon)\beta_Z}{\Bigl(\sigma_X+(1+\epsilon)(\alpha_X+\beta_Z)\Bigr)\sigma'_Z+(1+\epsilon)(\alpha_Z + 2\beta_X)},
\ee
and
\be
\tau^{new}_Z=\frac{\tau_Z+\epsilon (\alpha_Z+\beta_X) + (1+\epsilon)\beta_X   }{\Bigl(\sigma_X+(1+\epsilon)(\alpha_X+\beta_Z)\Bigr)\sigma'_Z+(1+\epsilon)(\alpha_Z + 2\beta_X)},
\ee
where
$\sigma'_Z={\rm max}(\frac{\sigma_Z+\beta_X}{\sigma_X+(1+\epsilon)(\alpha_X+\beta_Z)},1)$.

Further, for any code in first code family, one can construct a corresponding code in the second code family, with $K^{new}=K$ and $N^{new}$ polynomial in $N$, using an efficient randomized classical algorithm.

Finally, the same theorem holds with $O(\ldots),\Omega(\ldots)$ replaced with $O^*(\ldots)$ and $\Omega^*(\ldots)$ throughout, where $O^*(\ldots),\Omega^*(\ldots)$ denote
asymptotic behavior up to polylogs.
\end{theorem}

The construction of the code $C^{new}$ that we give below will be an existence proof that uses the Lovasz local lemma.  The efficient algorithm of theorem \ref{mainth} will follow
from the existence of efficient algorithms to find satisfying assignments guaranteed by the Lovasz local lemma\cite{mt}.
 In fact, by adjusting some of the constants hidden in the $O(\ldots)$ notation, we need only use the Locasz local lemma where there is a large gap between the parameters that we have and the worst possible parameters; i.e., the Lovasz local lemma guarantees that if there are ``bad events" that occur with probability $p$ and each event is independent of all but at most $d$ other events, then there is a nonzero probability that no bad events occur if $ep(d+1) \leq 1$.  By adjusting the constants, we can guarantee that we will be in the case that $ep(d+1)\leq C$, for any constant $C>0$.  We will not have a degree bound on the dependency graph in the application of the Lovasz local lemma, so we will not be able to use deterministic variants of the
algorithm of Ref.~\onlinecite{mt}.  However, it may be possible to use the algorithm of Ref.~\onlinecite{llldet} by adjusting constants here.

One application of these results is to the codes based on high dimensional tori of Ref.~\onlinecite{hdm}, which have
 $w_X,w_Z,q_X,q_Z=\Theta(\log(N))$ and which are conjectured to have distance $\Theta(N^{1-\epsilon})$.  If this conjecture holds, this would give a strongly LDPC quantum code family with distance $\Theta^*(N^{1-\epsilon})$ for any $\epsilon>0$.

Another application is to soundness of quantum codes.  While theorem \ref{mainth} does not imply anything about the soundness properties of a code, in section \ref{soundsec} we prove lemmas \ref{stabsplitlemmaeps},\ref{prodCeps} which show that for a code $C$ with  $w_X,w_Z,q_X,q_Z=\Theta(\log(N))$, the resulting
code $C^{new}$ has soundness parameters $\epsilon^{new}_X,\epsilon^{new}_Z$ which are at most polylogarithmically (in $N$) smaller than ${\rm min}(\epsilon_X,1),{\rm min}(\epsilon_Z,1)$ respectively.  Applied to other codes given in Ref.~\onlinecite{hdm}, based on a product of high-dimensional spheres, this allows the construction of strongly LDPC quantum code families with distance $\Theta^*(\sqrt{N})$ and inverse polylogarithmic $\epsilon_X,\epsilon_Z$.

Unfortunately, when applied to concatenated codes, the construction does not lead to a code with good distance properties.  Concatenated codes may have distance $\Theta(N)$ but at the cost of having a small number of generators which have weight $\Theta(N)$.  Consequently the resulting code $C^{new}$ will not have good distance.

However, one interesting application to concatenated codes is to reduce $q_X,q_Z$ without trying to reduce $w_X,W_Z$.  Typical concatenated codes will have $q_X,q_Z=\Theta(\log(N))$.  This can be done by using the construction in lemmas \ref{prodC},\ref{qsplitlemma2}.  This construction allows one to reduce $q_X,q_Z$ to $O(1)$ while only polylogarithmically increasing the number of physical qubits and without decreasing the distance.

A final application is on another conjectured code family.  The homological product code construction of
Ref.~\onlinecite{hp} was based on taking a homological product of two random codes; it was conjectured that
products of three or more codes would allow construction of codes with distance $\Theta(N)$ and generator weight
$O(N^{\alpha})$ (and also $q_X,q_Z$ which are $O(N^{\alpha})$ for $\alpha<1/2$.  For sufficiently small $\alpha$,
theorem \ref{mainth} implies that $C^{new}$ would have $\tau^{new}_X,\tau^{new}_Z>1/2$.  Indeed, in this case where $\alpha_X=\alpha_Z=\beta_X=\beta_Z=\alpha$, then we
have $$\tau_X^{new}=\tau_Z^{new}\;  \xrightarrow[\epsilon\rightarrow 0^+]\; \frac{1+\alpha}{1+6 \alpha},$$ and hence for $\alpha<1/4$, one can obtain $\tau_X^{new}=\tau_Z^{new}>1/2$.

In general, the quantum code $C^{new}$ may have $d_X \neq d_Z$.  However, in many circumstances, one is interested in the minimum distance ${\rm min}(d_X,d_Z)$.  In corollaries \ref{balB},\ref{balC}, we give a method of increasing whichever distance is smaller, at the cost of increasing the number of qubits.  
Corollary \ref{balC} shows that given a quantum code family which is strongly LDPC with $d_X d_Z=\Omega(N^{\mu})$, one can
construct a strongly LDPC quantum code family with ${\rm min}(\tilde d_X,\tilde d_Z)=\Omega(\tilde N^\nu)$ with
$$\nu=\frac{1}{3-\mu}.$$
Thus, for $\mu>1$, $\nu>1/2$.

\subsection{Codes from Chain Complexes}
We first review the relationship between CSS codes and chain complex\cite{csshomology1,csshomology2,csshomology3}.  This will be used later in the specific constructions of the paper.  We introduce it now because
in the next subsection we will review some ideas of Ref.~\onlinecite{bh} which motivate the construction here, and those ideas rely on the relationship.

Given a CSS quantum code, we construct a chain complex
$C_{q+1}\stackrel{\partial_{q+1}}{\rightarrow} C_q \stackrel{\partial_{q}}{\rightarrow} C_{q-1}$, for some integer $q$.  This notation means that $C_{q+1},C_q,C_{q-1}$ are vector spaces.  Since we are concerned with qubit quantum codes, they will be vector spaces over the field $\mathbb{F}_2$.  For qudit codes with prime dimension $p$, we can work over field $\mathbb{F}_p$.
The so-called ``boundary operator" $\partial_{q+1}$ is a linear map from $C_{q+1}$ to $C_q$ while the boundary operator $\partial_q$ is a linear map from $C_q$ to $C_{q-1}$.
The defining property of a chain complex is that ``the boundary of a boundary is zero", meaning in this case that
\be
\label{bbz}
\partial_q \partial_{q+1}=0.
\ee

The choice of the integer $q$ is simply a notational choice.  We say that ``the qubits are identified with the $q$-cells" to indicate the value of $q$.  However, in some cases, some choices of $q$ will be useful for certain topological or geometrical analogies.

The construction of the chain complex from the code is as follows.  The dimension of the vector space $C_q$ is equal to $N$, the number of qubits.
The dimension of $C_{q+1}$ is equal to $n_Z$, with each basis element in the standard basis corresponding to a distinct $Z$-type generator, and the dimension of $C_{q-1}$ is equal to $n_X$ with each basis element in the standard basis corresponding to a distinct $X$-type generator.  The matrix elements of the boundary operator $\partial_{q+1}$ are defined as follows: the element in the $i$-th row and $j$-th column is equal to $1$ if the $j$-th $Z$-type generator acts on qubit $i$, while it is zero otherwise.  The matrix elements of $\partial_q$ are defined similarly: the element in the $i$-th row and the $j$-th column is $1$ if the $i$-th $X$-type stabilizer acts on qubit $j$ and it is zero otherwise.  (This definition generalizes to qudit codes as follows: if a $Z$-type generator is of the form $Z_1^{a_1} Z_2^{a_2} \ldots Z_N^{a_N}$, then the corresponding column of $\partial_{q+1}$ has elements $a_1,a_2,\ldots$.)
Then, equation \ref{bbz} follows from the fact that the $Z$-type generators commute with the $X$-type generators.

Similarly, given a chain complex $C_{q+1}\stackrel{\partial_{q+1}}{\rightarrow} C_q \stackrel{\partial_{q}}{\rightarrow} C_{q-1}$ over $\mathbb{F}_2$, we
can define a CSS qubit quantum code by using $\partial_{q+1}$ to define the $Z$-type generators and using $\partial_{q}$ to define the $X$-type generators.
Note: often, given a manifold and a triangulation of the manifold, we may get a chain complex with more than $3$ vector spaces.  For example, given a $4$ manifold, we have $5$ different vector spaces, with chain complex $C_4 \stackrel{\partial_{4}}{\rightarrow} C_{3} \stackrel{\partial_{3}}{\rightarrow} C_2 \stackrel{\partial_{2}}{\rightarrow} C_{1} \stackrel{\partial_{1}}{\rightarrow} C_0$.  In this case, we pick some value of value of $q$ define the code and different choices of $q$ correspond to different codes.
Note: given a chain complex with only two vector spaces, $C_1,C_0$ we can define a classical code.

Given a vector $v \in C_q$, we say that the ``$Z$-type operator corresponding to $v$" is the product of Pauli $Z$ operators on qubits corresponding to nonvanishing entries of $v$, while the 
 ``$X$-type operator corresponding to $v$" is the product of Pauli $X$ operators on qubits corresponding to nonvanishing entries of $v$; in fact, more typical notation is also to introduce a vector space $C^q$ of so-called ``cochains" and let vectors in $C^q$ correspond to $X$-type operators and those in $C_q$ correspond to $Z$-type operators; here, we choose just to introduce one vector space.  Given a vector $v \in C_{q+1}$, we say that the ``corresponding $Z$-type stabilizer"  is the $Z$-type operator corresponding to $\partial_{q+1} v$ while given a
vector $v \in C_{q-1}$, we say that the ``corresponding $X$-type stabilizer is the $X$-type operator corresponding to $\partial_q^T v$.

\subsection{Geometric Interpretation and Comparison to Previous Work}
In Ref.~\onlinecite{bh}, a heuristic argument was given to suggest the existence of some kind of weight reduction procedure similar to that given here.  That argument relied on ideas from geometry and topology.  We now review the argument, as it motivates many of the specific constructions here.
However, the work in the present paper differs in several ways.  One difference is that here we give a specific algorithm, and give controlled estimates for the distance and number of qubits of the code $C^{new}$.  Also, the argument in Ref.~\onlinecite{bh} relied on a sequence of steps which used a manifold as an intermediary (first code $C$ was used to construct a cell complex, which was used to construct a manifold; then, a different triangulation of the manifold was given; then finally, that triangulation was used to construct a new code), while here all operations are expressed directly in terms of operations on the generators of a code, using the operations on the cell complex and manifold simply as motivation.  Finally, because we work directly on the generators of the code, rather than introducing a manifold, we can interchange $X$ and $Z$; so, for example, at one point we will define
a certain operation on $Z$ generators and use the ideas of Ref.~\onlinecite{bh} to motivate this operation geometrically in terms of subdividing simplices; then, we will apply the same operation to $X$ generators without giving a geometric interpretation in that case.  While a procedure acting directly on the generators was suggested in Ref.~\onlinecite{bh}, as we explain below some additional ideas are needed to make it work.

In the heuristic argument,
the CSS code $C$ is used to define a chain complex, with the qubits identified with the $2$-cells.
From this chain complex, we construct a cell complex.
There will be one $2$-cell for every qubit.
These $2$-cells will have the topology of a multiply-punctured sphere, where the number of punctures is equal to
the number of $X$-type generators acting on the given qubit.  The cell complex has one $1$-cell for each $X$-type generator; these $1$-cells all have the topology of a circle.
The boundary of a $2$-cell corresponding to a qubit is a sum of $1$-cells, with those $1$-cells
corresponding to the $X$-type generators acting on that qubit.  Thus, if there is some $X$-type stabilizer that is $X_1 X_2 X_3$, then there will be one $1$-cell corresponding to that generator, and this $1$-cell will be be attached to the three different $2$-cells which correspond to qubits $1,2,3$.

There will be one $3$-cell for every $Z$-type generator.  The boundary of this $3$-cell will be a sum of $2$-cells, corresponding to the qubits acted on by that generator.  Since the $X$-type and $Z$-type generators commute, this sum of $2$-cells has no boundary, so it is possible to take it as the boundary of some $3$-cell.

Then, in Ref.~\onlinecite{bh}, it was suggested that given this cell complex, one could embed it in general position in higher dimensions, and then suitably smooth thicken it to obtain a manifold with boundary; then one could attach this manifold to another copy of itself and identify the boundaries, to obtain a manifold without boundary.

Then, given this manifold, one can try to construct a cellulation with bounded local geometry, so that each cell attaches to only a bounded number of other cells.
One way to construct such a cellulation is by refining a cell.  For example, one can take a $3$-cell and split it into two different $3$-cells by adding a $2$-cell.  This has the effect of splitting a $Z$-type stabilizer into two different $Z$-type stabilizer, each of which act on fewer qubits.
To give an example of this in one lower dimension, consider a two dimensional toric code\cite{csshomology1,csshomology2}.  Suppose that one uses a cellulation of the torus which include some $2$-cell which is an $n$-gon for some large $n$.  Then, one can split this $n$-gon into two $(n/2)+1$-gons (assuming $n$ even; for $n$ odd one of the polygons after splitting will have one more edge than the other) by adding a $1$-cell cutting across that $n$-gon.

This procedure is used to construct the ``$Z$-type generator splitting step".  In subsection \ref{stabsplit} we explain this procedure in terms of its action on the generators of the code.  We describe it in a dual language: we use it to split $X$-type generators rather than splitting $Z$-type generators, so we call it an ``$X$-type generator splitting step".
This splitting step in terms of the generators in fact was already suggested in Ref.~\onlinecite{bh}.  In subsection \ref{stabsplit} we slightly modify how we apply it, by splitting each generator multiple times until the $X$-type generators all have weight $O(1)$.
Unfortunately, using just the $Z$-type and $X$-type generator splitting steps, as suggested in Ref.~\onlinecite{bh}, does not seem to lead to a procedure that works in general.
We need also a procedure to reduce to $q_X,q_Z$; without this additional procedure to reduce $q_X,q_Z$, the $Z$-type generator splitting steps reduces $w_Z$ but increases $w_X$ and vice-versa for the $X$-type generator steps, and the result does not lead to progress; since the increase in $w_X,w_Z$ depends on $q_Z,q_X$, adding this additional procedure makes the whole procedure work.

One way to reduce $q_X,q_Z$ is given in section \ref{altsplit}.  This way, however, also does not seem to suffice.  However, in subsection \ref{qsplit}, we give another way to reduce $q_X,q_Z$ which does suffice for our purposes.  This way in subsection \ref{qsplit} has a geometric interpretation, discussed later, which corresponds to the thickening of the cell complex.
These qubit splitting procedures were not in Ref.~\onlinecite{bh} and introducing them is the main new idea here.

\subsection{Overview}
The construction of $C^{new}$ will involve four steps, each of which takes some code as input and give some other code as output.  The code $C^{new}$ will result from the combined operation of all four steps.

The third and fourth step will be the dual ($X$ and $Z$ interchanged) of the first and second step, so the construction
and proofs for the third and fourth step will follow directly from those in the first and second step.

Notationally, we will describe the first and second step as procedures that take a code $C$ as input and output a code $\tilde C$.
The first two steps are described in 
section \ref{wxqz}, where we show how to generate a code $C'$ with parameters $w'_Z,q'_X$ both small by combining these two operations, using the output code $\tilde C$ from the first step as the input code $C$ to the second step, and then setting $C'$ equal to the output code $\tilde C$ from the second step.

We use a tilde to denote the parameters of $\tilde C$,
such as $\tilde w_X,\tilde w_Z,\tilde q_X,\tilde q_Z,\tilde d_X,\tilde d_Z,\tilde N$.
For these steps, the parameters will naturally be in pairs, $w_X,q_Z$ and $w_Z,q_X$, where parameters in a pair only have an effect on each other.  That is, the values of $\tilde w_X,\tilde q_Z$ will depend only on $w_X,q_Z$ and the values of $\tilde w_Z,\tilde q_X$ will depend only on $w_Z,q_X$.

The first step is the ``$X$-type generator splitting step" in
subsection \ref{stabsplit}, where we show how to produce a code $\tilde C$ with $\tilde w_X=O(1)$ from an arbitrary input code $C$.  The second step is the ``$Z$-type qubit splitting step" in subsection \ref{qsplit}, where we show how to produce a code
$\tilde C$ with $\tilde q_Z=O(1)$ from an arbitrary input code $C$, with this operation having the property that $\tilde w_X=w_X$.  Thus, code $C'$ has $w'_Z,q'_X=O(1)$.
Before giving the procedure in subsection \ref{qsplit}, we review the ``homological product" of codes in subsection \ref{homprod}.

The combined procedure in section \ref{wxqz} makes $w'_Z,q'_X$ both $O(1)$ and also has the property that parameters $w'_X,q'_Z$ remain $O(1)$ if both were $O(1)$ as input to the combined procedure.  Hence, by following this combined procedure with its dual (interchanging $X$ and $Z$) we succeed in producing a code $C^{new}$ from the original input code $C$
such that $C^{new}$ has parameters $w^{new}_X,w^{new}_Z,q^{new}_X,q^{new}_Z$ are $O(1)$.

After giving this construction and proving theorem \ref{mainth}, we show in section \ref{balD} how to increase whichever distance, $d_X$ or $d_Z$, of a code is smaller.   In section \ref{soundsec} we discuss soundness properties of the resulting code.
 In section \ref{altsplit} we give other ways of reducing $q_Z$.

\section{Reducing $w_X$ and $q_Z$}
\label{wxqz}
\subsection{Generator Splitting}
\label{stabsplit}
We now explain the $X$-type generator splitting step.
A geometric interpretation in terms of subdividing cells was given above for the dual operation, the $Z$-type generator splitting step.  The $X$-type splitting step can be also understood geometrically: if many $2$-cells $q_1,q_2,\ldots$ are attached to the same $1$-cell $e$, one can add an additional $2$-cell $c$ with the topology of a twice-punctured circle, with $1$-cells in its boundary labelled $e_1,e_2$.  Then, one can attach some of $2$-cells $q_1,\ldots$ to $e_1$ and the other ones to $e_2$.  This reduces the generators weight by roughly a factor of $2$; the procedure described here in fact corresponds to repeating that procedure many times until the generator weight is $O(1)$.

Let $S$ be an arbitrary $X$-type generator of the form $X_{q_1} X_{q_2} \ldots X_{q_w}$, where $q_1,\ldots,q_w$ are distinct qubits, with $w \geq 4$.  
Define a new code with $\tilde N=N+w-3$ qubits as follows.  The qubits of the new code are labelled $1,\ldots,N$ and $(m)$, for $1 \leq m \leq w-3$.  We refer to the qubits labelled $(m)$ as $m$ ``cut qubits".
The new code has the same $Z$-type generators as $C$ does and it has all of the $X$-type generators of $C$ other than generator $S$.  In addition, it has
$w-2$ $X$-type generators which are as follows:
$$X_{q_1} X_{q_2} X_{(1)}, \quad X_{(1)} X_{q_3} X_{(2)}, \quad X_{(2)} X_{q_4} X_{(3)}, \ldots, \quad X_{(w-4)} X_{q_{w-2}} X_{(w-3)}, \quad X_{(w-3)} X_{q_{w-1}} X_{q_w}.$$
For use below, let us define $T$ to be the set of these $w-2$ generators.
Thus, the new code has $\tilde n_X=n_X+(w-3)$ $X$-type generators.
(In the case of a qudit code, for each $(a)$, replace $X_{(a)}$ by $X_{(a)}^{-1}$ in either of the two generators acting on $(a)$.)

The new code will have $n_Z$ $Z$-type generators.
  For each $Z$-type generator $R$ of the old code, we define an $Z$-type generator $R'$ of the new code as follows.  Let $N$ be the set of integers $m$, for $1 \leq m \leq w-3 $ such that
$R$ anti-commutes with $X_{q_1} X_{q_2} \ldots X_{q_{m+1}}$.
Then let generator $R'$ be
\be
R'=R \prod_{a\in N} Z_{(a)}.
\ee
For use below, we say that $R$ is the generator in $C$ ``matching" to $R'$ and $R'$ is the generator of the new code ``matching" to $R$.
This completes the description of the new code after such a generator splitting step.
One may verify that all generators commute.

We define a new code $\tilde C$ by applying this generator splitting step to all $X$-type generators.  That is, we repeatedly apply this step to arbitrarily chosen generators until all generators have weight at most $3$ (since 
 all of the generators in $\tilde C$ that were not in the old code have weight at most $3$, they never need to be ``split again").
 We refer to qubits in $\tilde C$ as ``cut qubits" if they correspond to cut qubits in any of the splitting steps; other qubits are called ``non-cut qubits".
 
Thus,
\begin{lemma}
\label{stabsplitlemma}
The code $\tilde C$ has the following properties:
\begin{itemize}
\item[1.]  Let $\Delta$ be the sum over $X$ stabilizers of $w-3$, where $w$ is the weight of that stabilizer.
Then, $\tilde N=N+\Delta$ and $\tilde n_X=n_X+\Delta$ and $\tilde n_Z=n_Z$.
Since $\Delta\leq W_X$, we have $\tilde N \leq N + W_X$ and $\tilde n_X \leq n_X+W_X$.

\item[2.] $\tilde w_X \leq 3$.

\item[3.] $\tilde q_Z \leq {\rm max}(w_X q_Z/2,q_Z)$.

\item[4.] $\tilde w_Z \leq w_Z (q_X+1)$.

\item[5.] $\tilde q_X \leq {\rm max}(q_X,2)$.

\item[6.] $\tilde K = K$.

\item[7.] $\tilde d_X \geq d_X/(w_x/2+1)$.

\item[8.] $\tilde d_Z \geq d_Z$.

\item[9.] $\tilde W_X = W_X+\Delta \leq 2W_X$.

\item[10.] $\tilde W_Z \leq W_Z (q_X+1)$.
\end{itemize}
\begin{proof}
\begin{itemize}
\item[1.] In a single splitting step, the number of qubits increases by $w-3$ and the number of $X$ stabilizers increases by $w-3$, where $w \leq w_X$.  
The sum of $w-3$ over steps is equal to $\Delta$.

\item[2.] By construction.

\item[3.] Each cut qubit is in at most $w_X q_Z/2$ $Z$-type generators.  
To see this, suppose the cut qubit is labelled $(m)$ with $m+1 \leq w/2$.  Then, for each $R$, $m$ is in $N$ if
$R$ anti-commutes with $X_{q_1} X_{q_2} \ldots X_{q_{m+1}}$.  There are at most $q_Z w/2\leq q_Z w_X/2$ $Z$-type generators acting on those qubits $q_1,\ldots,q_{m+1}$.  If instead $m+1>w/2$, use the fact that
$m$ is in $N$ if 
$R$ anti-commutes with $X_{q_{m+2}} \ldots X_{q_w}$. For all non-cut qubits, the number of $Z$-type generators that they are in is unchanged by the splitting.

\item[4.] Each $Z$-type generator $R$ in $C$ of the form $Z_{q_1} \ldots Z_{q_a}$ matches to some $Z$-type generator in $\tilde C$; this generator is called $\tilde R$, with
$\tilde R$ equal to $R$ times some product of $Z$ operators on cut qubits.  In each step, where we split some $X$-type generator $S$, we may add more cut qubits to this product.  The number added is at most equal to the number of
qubits $q_1,\ldots,q_a$ which are in the support of $S$.  Hence, the total number of cut qubits in the support of $\tilde R$ is at most $a q_X \leq w_Z q_X$.

\item[5.] Each cut qubit in some step is in at most $2$ $X$-type generators.  For all other qubits, the number of $X$-type generators that they are in is unchanged by the splitting.

\item[6.]   For an arbitrary CSS quantum code with $N$ physical qubits, one can compute the number of logical qubits $K$ as follows.  Say that a set of generators are linearly independent if there is no nontrivial product of them (nontrivial meaning that it contains at least one such generator) which is equal to the identity operator.  Find a linearly independent set of $X$-type generators with the largest possible size; given generators $S_1,S_2,\ldots$, this can be done in a greedy fashion by starting with the set equal to $S_1$, and then adding each generator $S_i$ for $i>1$ to the set if this addition leaves the set linearly independent.  Suppose that there are $n^{ind}_X$ generators in the resulting set.  
In this case we say that ``$C$ has $n^{ind}_X$ linearly independent $X$-type generators".
Construct a similar set of linearly independent $Z$-type generators with $n^{ind}_Z$ generators.  Then, $K=N-n^{ind}_X-n^{ind}_Z$.

We will consider how these numbers $N,n^{ind}_X,n^{ind}_Z$ change when going from
a code $C$ to a new code after a single $X$-type generator splitting step.  We will show that the number of logical qubits is unchanged.  Since $\tilde C$ is obtained from $C$ after several such steps, it will follow that $\tilde K=K$.

The new code has $N+w-3$ qubits.  
We first show that the new code has $n^{ind}_X+w-3$ linearly independent $X$-type generators.
Let $I_X$ be a linearly independent set of $X$-type generators for code $C$ with $|I_X|=n^{ind}_X$.  Then, consider set
$$J_=I_X \cup \{X_{q_1} X_{q_2} X_{(1)},  X_{(1)} X_{q_3} X_{(2)},  X_{(2)} X_{q_4} X_{(3)}, \ldots,  X_{(w-4)} X_{q_{w-2}} X_{(w-3)}\}.$$
Note that the second set in the union above includes $w-3$ generators from the set $T$; in fact, we can choose any $w-3$ generators from $T$ in what follows.
 The set $J_X$ is linearly independent by construction (each of the generators $X_{q_1} X_{q_2} X_{(1)},  X_{(1)} X_{q_3} X_{(2)},  X_{(2)} X_{q_4} X_{(3)}, \ldots,  X_{(w-4)} X_{q_{w-2}} X_{(w-3)}$ is linearly independent from all others due to their action on qubits $(1),(2),\ldots$).  So, the new code has at least $n^{ind}_X+w-3$ linearly independent $X$-type generators.  Now, let $K_X$ be a maximal linearly independent set of $X$-type generators
for the new code.  This set must include at least $w-3$ generators from the set $T$; if not, the set $K_X$ would not be maximal.  Now, if $|K_X|>n^{ind}_X+w-3$, and if $K_X$ includes exactly $w-3$ generators from $T$, then if we consider
the set of generators in $K_X$ which are not in $T$, then this would give a linearly independent set of generators for $C$ with more than $n^{ind}_X$ generators in the set.  Similarly, if $K_X$ includes all $w-2$ generators from set $T$, then $(K_X \setminus T) \cup \{S\}$ would give a linearly independent set of generators for $C$ with more than $n^{ind}_X$ generators ($S$ must be linearly independent from $K_X \setminus T$ since $S$ is a product of generators in $T$).

Now we show that the new code has $n^{ind}_Z$ linearly independent $Z$-type generators.  Let $I_X$ be a maximal linearly independent set of $Z$-type generators for code $C$.
Consider the set of generators for the new code containing the generators matching to the generators in $I_X$; this gives a
linearly independent set of $Z$-type generators of the new code, so the new code has
at least $n^{ind}_Z$ linearly independent $Z$-type generators.
Similarly, given any linearly independent set of $Z$-type generators $J_X$ for the new code, consider the matching set of generators for $C$; i.e., simply remove any occurrences of operators $Z_{(m)}$ from the generators.  
We claim that the resulting set of generators for $C$ is linearly independent.  Indeed, if some product of the generators were equal to the identity, one may verify that all occurrences of operators $Z_{(m)}$ in the matching product of generators in $J_X$ would cancel out (if not, the matching product would not commute with all $X$-type generators of the new code), and so $J_X$ would not be linearly independent.

Hence, the new code has $N+w-3-(n^{ind}_X+w-3)-n^{ind}_Z=K$ logical qubits.

\item[7.]  Let $L$ be a logical $X$-type operator of $\tilde C$.  If $L$ does not act on the cut qubits, then it must commute with all stabilizers of $C$ and hence $L$ has weight at least
$d_X$.  Now, suppose that $L$ does act on the cut qubits and suppose that $L$ has weight $d$.  We will construct a logical operator $M$, with $M$ equal to $L$ multiplied by an element of the stabilizer group so that $M$ does not act on the cut qubits and so that $\wt(M) \leq (w_X/2+1) \wt(L)$.  Then, since $M$ has weight at least $d_X$, we have $\tilde d_X \geq d_X/(w_X/2+1)$.
To construct $M$, consider each cut qubit in $L$.  Suppose that
this cut qubit was labelled $(m)$ in some step splitting a stabilizer $S=X_{q_1} X_{q_2} \ldots X_{q_w}$.  Then ,if $m\leq w/2$, then multiply $L$ by the product of stabilizers
$(X_{q_1} X_{q_2} X_{(1)}) (X_{(1)} X_{q_3} X_{(2)}) (X_{(2)} X_{q_4} X_{(3)}) \ldots (X_{(m-1)} X_{q_{m+1}} X_{(m)})=X_{q_1} X_{q_2} \ldots X_{q_{m+1}} X_{(m)}$.  The resulting
operator does not act on cut qubit $(m)$ .  If $m>w/2$, then instead multiply $L$ by the product
$(X_{(m)} X_{q_{m+2}} X_{(m+1)}) (X_{(m+1)} X_{q_{m+3}} X_{(m_2)}) \ldots$.  This procedure replaces each cut
qubit by at most $w_X/2$ non-cut qubits.

\item[8.] We show that in each $X$-type stabilizer splitting step, the $Z$ distance is not decreased.  Let $L$ be a logical $Z$-type operator of the new code.  Note that the product of all stabilizers
in set $T$ is equal to stabilizer $S$.  Hence, $L$ must commute with all stabilizers of $C$ (that is, the stabilizers of $C$ act on qubits $1,\ldots,N$ and $L$ must commute with all of these
stabilizers as well as commuting with stabilizers acting on qubits $(1),(2),\ldots$).
Let $M$ be the product of the Pauli $Z$ operators in $L$ which act on qubits $1,\ldots,N$.  Either $M$ has weight at least $d_Z$ or else
$M$ is in the stabilizer group of $C$.  In the first case,
$L$ has weight at least $d_Z$, which is what we wish to show.  In the second case, since $M$ is a product of generators of $C$, by multiplying $L$ by the matching generators of the
new code, we obtain a logical operator for the new code which acts only on the cut qubits.  However, one may verify that the $Z$-type operator acting only on the cut qubits which commutes with all the generators is the identity operator.

\item[9.] In each splitting step, the total weight of $X$ stabilizers increases by $w-3$.

\item[10.] Same as proof of {\bf 4} above.
\end{itemize}

\end{proof}

\end{lemma}

\subsection{Codes from Chain Complexes and Homological Product}
\label{homprod}
In this section, we review the homological product of two quantum codes\cite{fh,bh,hp}.  The homological product we describe is slightly different from the one in Ref.~\onlinecite{bh}, which considered a ``single sector" version of this product for technical reasons.

Given two complexes, $C,C'$, their homological product, written $C \times C'$, is a complex defined as follows.
Let $D=C \times C'$.  Then,
\be
D_r = \oplus_{q} \Bigl( C_q \otimes C_{r-q} \Bigr)
\ee
and $D$ has boundary operator
\be
\label{prodbo}
\delta=\partial \otimes I + I \otimes \partial',
\ee
where $\partial,\partial'$ are the boundary operators of $C,C'$, with the appropriate subscript, and where $I$ is the identity operator.
More explicitly, inserting the subscript,
we have boundary operator
$$\delta_r=\oplus_q \Bigl(\partial_q \otimes I + I \otimes \partial'_{r-q}\Bigr)$$
acting on $D_r$.
That is, for example, in the case of $C$ being a complex
$C_{2}\stackrel{\partial_{2}}{\rightarrow} C_1 \stackrel{\partial_{1}}{\rightarrow} C_{0}$
and $C'$ being a similar complex, then
$\delta_2=(C_2 \otimes C'_0) \oplus (C_1\otimes C'_1) \oplus (C_0 \otimes C'_2)$.

This definition of the boundary operator on the product code is correct for complexes over $\mathbb{F}_2$.  For more general complexes, there is an extra sign:
$$\delta_r=\oplus_q (\partial_q \otimes I + (-1)^{q} I \otimes \partial'_{r-q}).$$
One may check that if $C,C'$ are complexes, then so is $D$, in that
\be
\delta_{r} \delta_{r+1}=0
\ee
for all $r$.

Using these two tools, we can define a homological product of two codes.  First, given the two codes, we construct chain complexes.  Then, we take the product of the two chain complexes.  Finally, we use the product complex to define a code.  There is some freedom in defining this code, as we need to pick a choice of $q$ to identify the qubits with the $q$-cells.

We can use results in algebraic topology to determine one important property of this code, namely the number of logical qubits.  Given
a complex
$C$, define $H_r(C)$ to be ${\rm ker}(\partial_{r})/{\rm im}(\partial_{r+1})$, where ${\rm ker},{\rm im}$ denote the kernel and image, respectively.
A vector in $C_r$ is said to ``represent nontrivial homology" if it is in the kernel of $\partial_r$ and not in the image of $\partial_{r+1}$.
The $Z$-type operator corresponding to such a vector is a $Z$-type logical operator.
Associate the qubits with $q$-cells.  Then\cite{review}:
\be
\label{Khom}
K={\rm dim}(H_q(C)).
\ee
Given a product of two complexes, we have
\be
H_r(C \times C')=\oplus_q \Bigl( H_q(C) \otimes H_{r-q}(C') \Bigr).
\ee
This is the so-called K\"{u}nneth formula; see Ref.~\onlinecite{hatcher}.
Let us write $b_r(C)={\rm dim}(H_r(C))$, so
that
\be
b_r(C\times C')=\sum_q b_q(C) b_{r-q}(C').
\ee

While the K\"{u}nneth formula enables us to determine the number of logical qubits of the product code from properties of the codes $C,C'$,
other properties of the product code cannot be determined so easily.  In Ref.~\onlinecite{bh}, statistical methods were used for products of random codes, while other results were given in Ref.~\onlinecite{hp}.

For use later, define $H^r(C)$ to be ${\rm ker}(\partial_{r+1}^T)/{\rm im}(\partial_r^T)$.We have ${\rm dim}(H^r(C))={\rm dim}(H_r(C))$.
A vector in $C_r$ is said to ``represent nontrivial cohomology" if it is in the kernel of $\partial^T_{r+1}$ and not in the image of $\partial^T_r$.
The $X$-type operator corresponding to such a vector is an $X$-type logical operator.

\subsection{$Z$-type Qubit splitting}
\label{qsplit}
We now explain the $Z$-type qubit splitting step.  This step is done in two substeps.  First, we define a code which consists of a homological product of code $C$ with a classical code $E$ corresponding to a one-dimensional chain complex; this chain complex is the cellulation of an interval.  Thus, this step may be geometrically interpreted as ``thickening" the complex, by taking its product with an interval.  The resulting code is called $D$.  Then, we observe that there is a large redundancy among the $Z$-type stabilizers of $D$: $Z$-type stabilizers at different values of the coordinate in the direction of the interval differ only by a product of other stabilizers.  This allows us to remove many of the stabilizers in $D$ and obtain a code $\tilde C$ with smaller $q_Z$.  Geometrically, one may view this as follows: since the complex has been thickened, we can attach the cells corresponding to stabilizers at different positions along the interval, to avoid attaching too many to any given cell.

We now explain this splitting step.
We give two separate explanations of the construction, one a more abstract
construction using the product formula, and then an explicit explanation of the code that results from this construction.
These two explanations leave certain choices undefined (the choice of the particular integers $k$ below).  We will show general
properties that hold for any choice of these integers in lemma \ref{prodC}.  Then, we will show how to choose these integers to reduce $q_Z$.

First, the abstract construction.
We define a splitting parameters $l$, which is an integer, $l>1$.
Given a quantum code, we define a chain complex $C$ as above, associating the qubits with the $2$-cells.
We define another chain complex $E_{1} \stackrel{\partial'_1}{\rightarrow} E_0$,
where ${\rm dim}(E_0)=l$ and ${\rm dim}(E_1)=l-1$, with
\be
\partial'_1=\begin{pmatrix}
1 & \\
1 & 1  \\
  & 1 & 1 \\
 & & & . & . \\
& & & & . & . \\
& & & & & . & . \\
& &  & & & & 1 & 1\\
& &  & & & &  & 1
\end{pmatrix}.
\ee
This chain complex can be interpreted geometrically as a cellulation of an interval, with $l$ $0$-cells and $(l-1)$ $1$-cells.
We have $b_1(E)=0,b_0(E)=1$.

We then define a product complex $D=C \times E$, and then define a code by associating the qubits with the $2$-cells.  This code will also be called $D$ and it should be clear in context whether we are referring to a code or to a complex.
Using the K\"{u}nneth formula, this code $D$ has the same number of logical qubits as $C$ does.
This code has $Nl+n_X(l-1)$ qubits and $ln_X$ $X$-type generators.

Finally, we also define a code $\tilde C$.  
This code $\tilde C$ will have the same number of qubits and $X$-type generators as $D$ does.
This code will be obtained by taking a subset of the $Z$-type generators of the generators of $D$, while taking all of the $X$-type generators.  What we will show is that although we take only a subset of the $Z$-type generators,
the code $\tilde C$ will have the same stabilizer group as $D$.

Recall that the $Z$-type generators are in one-to-one correspondence with basis elements of $D_3=C_3 \otimes E_0 \oplus C_2 \otimes E_1$.  We keep all $Z$-type generators corresponding to basis elements of $C_2 \otimes E_1$.  However, for basis elements of $C_3$,
we keep only one $Z$-type generator.  Let $w_1,\ldots,w_l$ be basis vectors for $E_0$ in the standard basis.
For each basis element $v\in C_3$ in the standard basis, we pick one integer $k$, with $1 \leq k \leq l$, and we keep the generator corresponding to $v \otimes w_k$.

Now, the explicit construction.
Given a code with $N$ qubits, we define a new code with $\tilde N=Nl+n_X(l-1)$ qubits.
The qubits in $C$ are labelled by $q=1,\ldots,N$.  Some of the qubits in $\tilde C$ are labelled by a pair, $(q,k)$, with $1\leq q \leq N$ and $1 \leq k \leq l$.
The remaining $n_X(l-1)$ qubits in $\tilde C$ are labelled by a pair $[s_X,k]$ for $1 \leq s_X \leq n_X$ and $1 \leq k \leq l-1$.  Note that we use parenthesis $(\ldots,\ldots)$ to label some of the qubits in $\tilde C$ and brackets $[\ldots,\ldots]$ to label other ones.
If there are $n_X$ $X$-type generators in $C$, then there are $\tilde n_X=ln_X$ $X$-type generators in $\tilde C$.
There will be $\tilde n_Z=n_Z+(l-1)N$ $Z$-type generators in $\tilde C$.

For each $X$-type generator labelled by $s_X$ with $1 \leq s_X \leq n_X$, if the generator has the form $X_{q_1} X_{q_2} \ldots X_{q_w}$ in $C$,
where $1 \leq q_1,\ldots,q_w \leq N$, we define $l$ different $X$-type generators in $\tilde C$, of the form
$$X_{(q_1,k)} X_{(q_2,k)} \ldots X_{(q_w,k)} X_{[s_X,k-1]} X_{[s_X,k]}$$ for $k=2,\ldots,l-1$ and
$$X_{(q_1,k)} X_{(q_2,k)} \ldots X_{(q_w,k)} X_{[s_X,k]}$$ for $k=1$
and
$$X_{(q_1,k)} X_{(q_2,k)} \ldots X_{(q_w,k)} X_{[s_X,k-1]}$$ for $k=l$.

For each $Z$-type generator of the form $Z_{q_1} Z_{q_2} \ldots Z_{q_w}$ in $C$,
where $1 \leq q_1,\ldots,q_w \leq N$, we define one $Z$-type generator in $\tilde C$, of the form
$$Z_{(q_1,k)} Z_{(q_2,k)} \ldots Z_{(q_w,k)}$$ for some value of $k$ with $1\leq k  \leq l$.  Different $Z$-type generators in $C$
 may have different values of $k$ when constructing the corresponding $Z$-type generator in $\tilde C$ (later we explain how we choose these values of $k$ to improve the parameter $q_Z$ in $\tilde C$; however, many of the properties of the new code will be independent of the choices of $k$ so we leave it unspecified for now).
In addition, for each qubit $q$ in $C$, we define $(l-1)$ $Z$-type generators in $\tilde C$.
These generators are of the form $$Z_{(q,k)} Z_{(q,k+1)} \prod_{s_X \ni q} Z_{[s_X,k]}$$ for $1 \leq k \leq t-1$.

\begin{lemma}
\label{prodC}
The code $\tilde C$ has the following properties:
\begin{itemize}
\item[1.] All generators commute with each other.  $\tilde N= Nl+n_X(l-1)$ and  $\tilde n_Z=n_Z+(l-1)N$ and $\tilde n_X=ln_X$.  

\item[2.] Codes $\tilde C$ and $D$ have the same stabilizer group.
Explicitly, for each $Z$-type generator of the form $Z_{q_1} Z_{q_2} \ldots Z_{q_w}$ in $C$,
the stabilizer group  of $\tilde C$ contains
$Z_{(q_1,m)} Z_{(q_2,m)} \ldots Z_{(q_w,m)}$ for all $1 \leq m \leq l$.

\item[3.] $\tilde  K =K$.

\item[4.] $\tilde w_X=w_X+2$ if $l\geq 3$ and $\tilde w_X=w_X+1$ if $l=2$.

\item[5.] $\tilde w_Z={\rm max}(w_Z,2+q_X)$.

\item[6.] $\tilde q_X={\rm max}(q_X,2)$.

\item[7.] $\tilde d_X=l d_X$.

\item[8.] $\tilde d_Z=d_Z$.

\item[9.] $\tilde W_X \leq l(W_X+2n_X)$.

\item[10.] $\tilde W_Z\leq W_Z+2(l-1)N+(l-1)W_X$.
\end{itemize}
\begin{proof}
\begin{itemize}
\item[1.] From the abstract construction the commutativity follows immediately: $D$ is a chain complex, so the generators of code $D$ commute, and $\tilde C$ has only a subset of the generators.  In the explicit construction, one can check this directly.

\item[2.] From the abstract construction, let $v$ be some basis element of $C_3$ in the standard basis.
Let $\tilde C$ include the generator corresponding to basis element $v \otimes w_k$.  Let $m$ be some integer, $m \neq k$, with $1 \leq m \leq l$.
Let $u$ be a vector in $E_1$ with $\partial'_1 u=w_k+w_m$; such a $u$
of the form $(0,\ldots,0,1,\ldots,1,0,\ldots,0)$.  Then, $v \otimes u \in D_4$ and $\delta_4 (v \otimes u)=(\partial v) \otimes u + v \otimes w_k + v \otimes w_m$.  Since $\delta_3 \delta_4=0$, the stabilizer corresponding to $\delta_4 (v\otimes u)$ is equal to the identity operator (intuitively, the image of $\delta_4$ gives some redundancies among stabilizers).
The vector $(\partial v) \otimes u$ is in $C_2 \otimes C_1$ so that the corresponding stabilizers are in $\tilde C$.  Hence,
the element of the stabilizer group of $D$ corresponding to $v \otimes w_k + v \otimes w_m$ is  in the stabilizer group of $\tilde C$.  Hence, since the stabilizer corresponding to $v \otimes w_k$ is in the
stabilizer group of $\tilde C$, so is the stabilizer of $D$ corresponding to $v \otimes w_m$.
Explicitly, one may check this by multiplying as follows: assume $m>k$.  Multiply $Z_{(q_1,k)} Z_{(q_2,k)} \ldots Z_{(q_w,k)}$ by
$$\prod_{i=k}^{m-1} \prod_{a=1}^w \Bigl( Z_{(q_a,i)} Z_{(q_a,i+1)} \prod_{s_X \ni q_a} Z_{[s_X,i]} \Bigr).$$
Then, using the commutativity of the generators of the code $C$, one may verify that the terms of the form $Z_{[s_x,k]}$ disappear.

\item[3.]  This follows from the K\"{u}nneth formula for $D$ and from the fact that $\tilde C,D$ have the same stabilizer group.

\item[4.] Follows directly from the given $X$-generators.

\item[5.] Follows directly from the given $Z$-generator.  The generators of form $Z_{(q_1,k)} Z_{(q_2,k)} \ldots Z_{(q_w,k)}$ have maximum weight $w_Z$ while those
of form $Z_{(q,k)} Z_{(q,k+1)} \prod_{s_X \ni q} Z_{[s_X,k]}$ have maximum weight $2+q_X$.

\item[6.]  The number of generators acting on qubits $(q,k)$ has maximum $q_X$, while the number acting on qubits $[s_X,k]$ has maximum $2$.

\item[7.]   First, note that $\tilde d_X \leq l d_X$ because given any $X$-type logical operator for code $C$, with $L=X_{q_1} X_{q_2} \ldots X_{q_{a}}$, where $a \geq d_X$, the operator
$M=\prod_{k=1}^l \Bigl( X_{(q_1,k)} X_{(q_2,k)}
\ldots X_{(q_{a},k)}\Bigr)$ is a logical operator for $\tilde C$. 
The K\"{u}nneth formula can be used to show that this is a logical operator.
Indeed, if operator $L$ is the operator corresponding to vector $v \in C_2$, and $w=\sum_{k=1}^l w_k$, where $w_k$ are the standard basic vectors for $E_0$,
then $M$ is the operator corresponding to $v \otimes w$ and since $v,w$ both represent nontrivial cohomology, so do $v \otimes w$ (this is a further result also known as K\"{u}nneth).

Next, we show that $\tilde d_X \geq l d_X$.
In the above paragraph, we constructed logical operators corresponding to vectors $v \otimes w$, where $v$ represents nontrivial cohomology and $w=\sum_{k=1}^k w_k$.
By K\"u{nneth}, every logical operator can be written as an operator corresponding to such a vector $v \otimes w + \delta_2^T u$, where $v,w$ are as in the above paragraph and $u$ is arbitrary.  Let $u=\sum_{k=1}^l u_k \otimes w_k$, with $u_k \in C_1$ and $w_k$ standard basis elements of $E_0$.
Consider the projection of $v \otimes w + \partial_2^T u$ into $C_2 \otimes E_0$.  This is equal to $\sum_{k=1}^l (v+\partial_2^T u_k) \otimes w_k$.  Each vector $v+\partial_2^T u_k$ is a logical operator for $C$ and hence has weight at least $d_X$, so $\wt(v\otimes w + \delta_2^T u)\geq l d_X$.

\item[8.]  First, note that $\tilde d_Z \leq d_Z$ because given any $Z$-type logical operator $L$ for code $C$, with $L=Z_{q_1} Z_{q_2} \ldots Z_{q_{d_Z}}$, the operator
$M=Z_{(q_1,k)} Z_{(q_2,k)} \ldots Z_{(q_{d_Z},k)}$ is a logical operator for $\tilde C$, for any $k$, $1\leq k \leq l$.  The K\"{u}nneth formula can be used to show that this is a logical operator.
Indeed, if operator $L$ is the operator corresponding to vector $v \in C_2$, and $w_k$ is one of the standard basic vectors for $E_0$
then $M$ is the operator corresponding to $v \otimes w_k$ and since $v,w_k$ both represent nontrivial homology, so do $v \otimes w_k$.

Next, we show that $\tilde d_Z \geq d_Z$.  Let $L$ be a $Z$-type logical operator for $\tilde C$.  By multiplying $L$ by a product of generators of the form $Z_{(q,k)} Z_{(q,k+1)} \prod_{s_X \ni q} Z_{[s_X,k]}$, we can construct an operator $M$ that does not act on any qubit $(q,k)$ for $k>1$.  To see this, do it iteratively.  Set $M=L$ initially.  Then, if an operator $Z_{(q,k+1)}$ appears in $M$, multiply by the generator above, and replace $M$ by the resulting operator, continuing until no such operators are left.  Thus, $M$ is a product of operators $Z_{(q,1)}$ as well as operators $Z_{[s_X,k]}$.  Further, the number of operators of the form $Z_{(q,1)}$ in $M$ is upper bounded by the number of operators $Z_{(q,k)}$ in $L$ which is upper bounded by the weight of $L$.
Recall that for each stabilizer $s_X=X_{q_1} \ldots X_{q_w}$ in $C$, the code $\tilde C$ has $X$-type stabilizers $X_{(q_1,k)} X_{(q_2,k)} \ldots X_{(q_w,k)} X_{[s_X,k-1]} X_{[s_X,k]}$ for $k=2,\ldots,l-1$ and $X_{(q_1,k)} X_{(q_2,k)} \ldots X_{(q_w,k)} X_{[s_X,k-1]}$ for $k=l$.  The operator $M$ must commute with all these stabilizers and hence must commute with 
$X_{[s_X,k-1]} X_{[s_X,k]}$ for $k=2,\ldots,l-1$ and $X_{[s_X,k-1]}$ for $k=l$.   This means that it must not act on qubit $[s_X,k-1]$ (otherwise it would not commute with the last such stabilizer); hence it must not act on qubit $[s_X,k-2]$ (else it would not commute with the stabilizer for $k=l-1$).  Proceeding in this fashion, $M$ must not act on any qubits of form $[s_X,k]$
and so $M$ acts only on qubits of form $(q,1)$.  However, then since $M$ commutes with all stabilizers 
$X_{(q_1,1)} X_{(q_2,1)} \ldots X_{(q_w,k)} X_{[s_X,1]}$, it follows that $M$ commutes with 
$X_{(q_1,1)} X_{(q_2,1)} \ldots X_{(q_w,1)}$.
So, if $M$ is a product $M=Z_{(r_1,1)} Z_{(r_2,1)} \ldots$ for some sequence $r_1,r_2,\ldots,$ then the product $Z_{r_1} Z_{r_2} \ldots$ commutes with all stabilizers in $C$.  Further, this product $Z_{r_1} Z_{r_2}$ must be a logical operator as if it were a product of $Z$-type stabilizers of $C$ then $M$ would be a product of $Z$-type stabilizers of $\tilde C$ (abstractly, we have shown that the vector corresponding to $M$ is of the form $v \otimes w_1$ and so $v$ must represent nontrivial homology).  So, the weight of $M$ must at least equal $d_Z$.

\item[9.] Immediate from the form of the $X$ generators.

\item[10.] The sum of weights of $Z$-type generators of the form
$Z_{(q_1,k)} Z_{(q_2,k)} \ldots Z_{(q_w,k)}$ is equal to $W_Z$.
The sum of weights of generators of the form $Z_{(q,k)} Z_{(q,k+1)} \prod_{s_X \ni q} Z_{[s_X,k]}$ is equal to $2(l-1)N$ plus $(l-1)$ times the sum over qubits of the number of $X$-type
stabilizers acting on that qubit.  However, the sum over qubits of the number of $X$-type
stabilizers acting on that qubit is equal to $W_X$.
\end{itemize}
\end{proof}
\end{lemma}

The value of $\tilde q_Z$ will depend on which generators $Z_{(q_1,k)} Z_{(q_2,k)} \ldots Z_{(q_w,k)}$  we choose to include.  If we keep the same $k$ for all generators, then we have no improvement in $\tilde q_Z$.  However, if we choose different $k$ for different generators then we can reduce $\tilde q_Z$ by making different generators act on different qubits $(q,k)$.  We now show the existence of a choice of $\tilde q_Z$ with certain properties:
\begin{lemma}
\label{qsplitlemma2}
Let $w$ be any positive integer.
For $l$ sufficiently large that
\be
2e {q_Z \choose w+1} (\frac{1}{l})^{w+1} {\rm min}(q_Z w_Z,N) l \leq 1,
\ee
there is a choice of $k$ for each $Z$-type generator such that
\be
\tilde q_Z \leq {\rm max}(w+2,w_X).
\ee
\begin{proof}
For each $Z$-type generator, choose $k$ independently and uniformly from $1,\ldots,l$.  We show that this gives the desired $\tilde q_Z$ with positive probability.

Consider a given qubit $q$.  There are at most $q_Z$ $Z$-type generators acting on that qubit in code $C$.
We now estimate the probability distribution of the maximum number of $Z$-type generators acting on a qubit of the form $(q,m)$ in code $\tilde C$ for given $q,m$.
This is equivalent to the ``balls into bins" problem: we have $l$ different bins (the different values of $m$, for $m=1,\ldots,l$).  We drop at most $q_Z$ balls into these bins, dropping each ball independently into a bin chosen uniformly (the balls are independent because each one corresponds to a different generator).  We then wish to determine the probability distribution of the maximum number of balls in a given bin.
The balls into bins problem is well-studied and one usually instead considers the maximum of this number over bins.
However, we will give a slightly different derivation because of the particular parameter regime that we consider: we are interested in $l>q_Z$ so that the number of balls in a given bin is likely to be small.

Consider a given bin (i.e., a given value of $m$).  Let us assume that $l \geq q_Z$.  If there are $q_Z$ $Z$-type generators acting on the given qubit, then probability of having more than $w$ balls in that bin is
\begin{eqnarray}
&&\sum_{u> w, u \leq q_Z} {q_Z \choose u} (\frac{1}{l})^u (1-\frac{1}{l})^{q_Z-u}\\ \nonumber
& \leq & {q_Z \choose w+1} (\frac{1}{l})^{w+1} \sum_{u>w, u\leq q_Z} \frac{(w+1)!}{u!} \frac{(q_z-w-1)!}{(q_z-u)!}
(\frac{1}{l})^{u-w-1} \\ \nonumber
& \leq & {q_Z \choose w+1} (\frac{1}{l})^{w+1} \sum_{u>w, u\leq q_Z} \frac{(w+1)!}{u!}
(\frac{q_Z}{l})^{u-w-1}  \\ \nonumber
& \leq & 2{q_Z \choose w+1} (\frac{1}{l})^{w+1},
\end{eqnarray}
where we used that $l \geq q_Z$ and $w\geq 1$ (in fact, a slightly tighter bound holds, since for $w=1$, we get $2!(1/2!+1/3!+1/4!+\ldots)=2(e-2)$.
If there are fewer than $q_Z$ $Z$-type generators acting on that qubit, then the probability is smaller than this.

There at $Nl$ choices of $(q,m)$.  We now use the Lovasz local lemma.  For each pair $(q,m)$, we define the event that that event has more than $w$ $Z$-type generators acting on it in code $\tilde C$; i.e., the probability of having more than $w$ balls in the corresponding bin.  This event is independent of all but at most
${\rm min}(q_Z w_Z,N) l$ events for other choices $(q',m')$.  To see this, note that each qubit $q$ has at most $q_Z$ generators acting on it, each of which act on at most $w_Z$ qubits.
Hence, by the Locasz local lemma,
if
$$2e {q_Z \choose w+1} (\frac{1}{l})^{w+1} {\rm min}(q_Z w_Z,N) l \leq 1,$$
then there is a nonzero probability that none of the events occurs.

If none of the events occurs, then for each pair $(q,m)$ there are at most $w$ generators of the form $Z_{(q_1,k)} Z_{(q_2,k)} \ldots Z_{(q_w,k)}$
acting on the qubit with that label.   However, there are also two (or one if $l=2$) generators of the 
 $Z_{(q,k)} Z_{(q,k+1)} \prod_{s_X \ni q} Z_{[s_X,k]}$ acting on the given qubit.  Hence, there are at most $w+2$ generators acting on the given qubit.
Considering qubits labelled $[s_X,k]$, there are at most $w_X$ generators of the form
$Z_{(q,k)} Z_{(q,k+1)} \prod_{s_X \ni q} Z_{[s_X,k]}$ acting on the given qubit.
\end{proof}
\end{lemma}

\subsection{Combined Effect}
\label{combeffect}
Let $C'$ be the code resulting from first applying the $X$-type generator splitting step and then the $Z$-type qubit splitting step.

Then,
\begin{lemma}
\label{comblemma}
Let $w,l$ be chosen such that
\be
\label{obey}
2e {w_X q_Z \choose w+1} (\frac{1}{l})^{w+1} {\rm min}(w_X q_Z/2,N+W_X) l \leq 1,
\ee
The code $C'$ has the following properties.
\begin{itemize}
\item[1.] $K'=K$ and $N' \leq (N+W_X)l+(n_X+W_X)(l-1)\leq l(N+n_X+2W_X)$ and $n'_X\leq l (n_X+W_X)$ and $n'_Z\leq n_Z+(l-1) (N+W_X)$.

\item[2.] $w'_X \leq 5$, and $q'_Z \leq {\rm max}(w+2,3)$.

\item[3.] $w'_Z \leq {\rm max}(w_Z(q_X+1),2+q_X)$ and $q'_X \leq q_X$.

\item[4.] $d'_Z \geq d_Z$ and $d'_X \geq l d_X/(w_X/2+1)$.

\item[5.] $W'_Z \leq W_Z(q_X+1)+2(l-1)N+4(l-1)W_X$ and $W'_X \leq 4lW_X+2ln_X$.
\end{itemize}
\begin{proof}
This follows from lemmas \ref{stabsplitlemma},\ref{prodC},\ref{qsplitlemma2}.
\end{proof}
\end{lemma}

We restate this in terms of code families using $O(\ldots)$ notation:
\begin{lemma}
\label{clasympt}
Let $\epsilon$ be a positive constant.
Consider a family of quantum codes, with
$w_X=O(N^{\alpha_X}),w_Z=O(N^{\alpha_Z}),q_X=O(N^{\beta_X}),q_Z=O(N^{\beta_Z}),W_X=O(N^{\sigma_X}),W_Z=O(N^{\sigma_Z}),d_X=\Omega(N^{\tau_X}),d_Z=\Omega(N^{\tau_Z})$.
Assume that $\sigma_X,\sigma_Z\geq 1$.
Then, there exists a choice $w,l$ obeying Eq.~(\ref{obey}) such that the code family $C'$ has 
$w'_X=O(1),w'_Z=O((N')^{\alpha'_Z}),q'_X=O((N')^{\beta'_X}),q'_Z=O(1),W'_X=O((N')^{\sigma'_X}),W'_Z=O((N')^{\sigma'_Z}),d'_X=\Omega((N')^{\tau'_X}),d'_Z=\Omega((N')^{\tau'_Z})$ and $K'=K$ and $N'$ polynomial in $N$, with
\begin{itemize}
\item[1.] $\alpha'_Z=\frac{\alpha_Z+\beta_X}{\sigma_X+(1+\epsilon)(\alpha_X+\beta_Z)}$.

\item[2.] $\beta'_X=\frac{\beta_X}{\sigma_X+(1+\epsilon)(\alpha_X+\beta_Z)}$.

\item[3.] $\sigma'_X=1$.

\item[4.] $\sigma'_Z={\rm max}(\frac{\sigma_Z+\beta_X}{\sigma_X+(1+\epsilon)(\alpha_X+\beta_Z)},1)$.

\item[5.] $\tau'_X=\frac{\tau_X+\epsilon\alpha_X+(1+\epsilon)\beta_Z}{\sigma_X+(1+\epsilon)(\alpha_X+\beta_Z)}$.

\item[6.] $\tau'_Z=\frac{\tau_Z}{\sigma_X+(1+\epsilon)(\alpha_X+\beta_Z)}$.
\end{itemize}

Further, the same result holds with $O(\ldots),\Omega(\ldots)$ replaced with $O^*(\ldots)$ and $\Omega^*(\ldots)$ throughout.

\begin{proof}

Choose
\be
\label{lchoicehere}
l=\lceil (w_X q_Z)^{1+\epsilon}\rceil.
\ee
Then, 
$2e {w_X q_Z \choose w+1} (\frac{1}{l})^{w+1} {\rm min}(w_X q_Z/2,N+W_X) l \leq
e (w_X q_Z/l)^{w+1} w_X q_Z l=e (w_X q_Z)^{w+2} l^{-w}\leq e (w_X q_Z)^{w+2} (w_X q_Z)^{-(1+\epsilon)w}$.
Now choose
\be
\label{wchoicehere}
w=\lceil 2/\epsilon \rceil+1.
\ee
Then, $e (w_X q_Z)^{w+2} (w_X q_Z)^{-(1+\epsilon)w} \leq e (w_X q_Z)^{-1}$ which is $\leq 1$ since we can assume without loss of generality that
$w_X q_Z \geq 3$.

Hence, $l=O(N^{(1+\epsilon)(\alpha_X+\beta_Z)})$.
Using these values of $w,l$ in lemma \ref{comblemma}, we find that
\begin{itemize}
\item[1.] $N'=O(N^{\sigma_X+(1+\epsilon)(\alpha_X+ \beta_Z)})$ and $n'_X=O(N^{\sigma_X+(1+\epsilon)(\alpha_X +\beta_Z)})$ and $n'_Z=O(N^{\sigma_X+(1+\epsilon)(\alpha_X +\beta_Z)})$.

\item[2.]  $w'_X=O(1),q'_Z=O(1)$.

\item[3.] $w'_Z = O(N^{\alpha_Z+\beta_X})$ and $q'_X=O(N^{\beta_X})$.

\item[4.] $d'_Z =\Omega(N^{\tau_Z})$ and $d'_X =\Omega(N^{\tau_X+\epsilon \alpha_X + (1+\epsilon) \beta_Z})$.

\item[5.] $W'_Z = O(N^{\sigma_Z + \beta_X})+O(N')$ and $W'_X = O(N^{\sigma_X+(1+\epsilon)(\alpha_X+ \beta_Z)})$.
\end{itemize}

Re-expressing these asymptotic values in terms of $N'$, we obtain the scaling above.
\end{proof}
\end{lemma}

\section{Dual Splitting and Combined Effect: Proof of Main Theorem}
We can now prove theorem \ref{mainth}, which we restate here:
\begin{unth}
Let $\epsilon$ be any positive constant.
Consider a family of quantum codes, with
$w_X=O(N^{\alpha_X}),w_Z=O(N^{\alpha_Z}),q_X=O(N^{\beta_X}),q_Z=O(N^{\beta_Z}),W_X=O(N^{\sigma_X}),W_Z=O(N^{\sigma_Z}),d_X=\Omega(N^{\tau_X}),d_Z=\Omega(N^{\tau_Z})$.
Assume that $\sigma_X,\sigma_Z\geq 1$.
Then, there exists a quantum code family $C^{new}$ which is strongly LDPC with 
$d^{new}_X=\Omega((N^{new})^{\tau^{new}_X}),d^{new}_Z=\Omega((N^{new})^{\tau^{new}_Z})$, with
\be
\tau^{new}_X=\frac{\tau_X+\epsilon\alpha_X+(1+\epsilon)\beta_Z}{\Bigl(\sigma_X+(1+\epsilon)(\alpha_X+\beta_Z)\Bigr)\sigma'_Z+(1+\epsilon)(\alpha_Z + 2\beta_X)},
\ee
and
\be
\tau^{new}_Z=\frac{\tau_Z+\epsilon (\alpha_Z+\beta_X) + (1+\epsilon)\beta_X   }{\Bigl(\sigma_X+(1+\epsilon)(\alpha_X+\beta_Z)\Bigr)\sigma'_Z+(1+\epsilon)(\alpha_Z + 2\beta_X)},
\ee
where
$\sigma'_Z={\rm max}(\frac{\sigma_Z+\beta_X}{\sigma_X+(1+\epsilon)(\alpha_X+\beta_Z)},1)$.

Further, for any code in first code family, one can construct a corresponding code in the second code family, with $K^{new}=K$ and $N^{new}$ polynomial in $N$, using an efficient randomized classical algorithm.

Finally, the same theorem holds with $O(\ldots),\Omega(\ldots)$ replaced with $O^*(\ldots)$ and $\Omega^*(\ldots)$ throughout, where $O^*(\ldots),\Omega^*(\ldots)$ denote
asymptotic behavior up to polylogs.
\begin{proof}
This follows from applying lemma \ref{clasympt}, interchanging $X$ and $Z$, re-applying lemma \ref{clasympt}, and re-interchanging $X$ and $Z$.  One finds
\be
\tau^{new}_X=\frac{\tau_X+\epsilon\alpha_X+(1+\epsilon)\beta_Z}{\sigma_X+(1+\epsilon)(\alpha_X+\beta_Z)} \frac{1}{\sigma'_Z+(1+\epsilon)(\alpha'_Z + \beta'_X)},
\ee
and
\be
\tau^{new}_Z=\frac{    \frac{\tau_Z}{\sigma_X+(1+\epsilon)(\alpha_X+\beta_Z)}+\epsilon \alpha'_Z + (1+\epsilon)\beta'_X   }
{ \sigma'_Z+(1+\epsilon)(\alpha'_Z +\beta'_X)   },
\ee
where
$\alpha'_Z=\frac{\alpha_Z+\beta_X}{\sigma_X+(1+\epsilon)(\alpha_X+\beta_Z)}$ and
$\beta'_X=\frac{\beta_X}{\sigma_X+(1+\epsilon)(\alpha_X+\beta_Z)}$ and
$\sigma'_Z={\rm max}(\frac{\sigma_Z+\beta_X}{\sigma_X+(1+\epsilon)(\alpha_X+\beta_Z)},1)$.
After some simplifying algebra, one gets the result above.
\end{proof}
\end{unth}

\section{Balancing Distance}
\label{balD}
We now give a corollary of lemma \ref{prodC} which can be used to increase $d_X$ or $d_Z$ at the cost of increasing $N$.
This is useful because it allows us to increase whichever distance is smaller, so that for the new code $d_X=d_Z$.  We call this ``balancing the code".

\begin{corollary}
\label{balB}
Suppose that code $C$ is strongly LDPC.  Then, applying the construction of lemma \ref{prodC}, the resulting code $\tilde C$ is strongly LDPC and
\begin{itemize}
\item[1.] $\tilde N=O(Nl)$.

\item[2.] $\tilde  K =K$.

\item[3.] $\tilde d_X=l d_X$.

\item[4.] $\tilde d_Z=d_Z$.
\end{itemize}

Thus, choosing $l\approx d_Z/d_X$, we obtain a code with $\tilde d_X=\tilde d_Z$ and $\tilde N=O(N d_Z/d_X)$.
\begin{proof}
Corollary of lemma \ref{prodC}.  Since $q_X$ is $O(1)$, $n_X=O(N)$, so $\tilde N=O(Nl)$.  The value of $\tilde q_Z$ holds for any choice of $k$ for the $Z$-type stabilizers; indeed, the code $D$ has $\tilde q_Z=O(1)$.
\end{proof}
\end{corollary}

This corollary has the following application to the question of finding quantum code families with $q_X,q_Z,w_X,w_Z$ all $O(1)$ and with minimum distance ${\rm min}(d_X,d_Z)=\Omega(N^\nu)$ for some $\nu>1/2$.  

\begin{corollary}
\label{balC}
Assume that there exists a code family which is strongly LDPC and with $d_X d_Z=\Omega(N^{\mu})$.  Then, applying the construction of corollary \ref{balB} (or its dual with $X,Z$ interchanged), the resulting code family is strongly LDPC and with ${\rm min}(\tilde d_X,\tilde d_Z)=\Omega(\tilde N^\nu)$ with
$$\nu=\frac{1}{3-\mu}.$$
Thus, for $\mu>1$, $\nu>1/2$.
\begin{proof}
Assume without loss of generality note that $\tilde C$ has $d_X \leq d_Z$.  Choosing $l=\lceil d_Z/d_X \rceil$, then $\tilde C$ has $\tilde d={\rm min}(\tilde d_X,\tilde d_Z)=d_Z$ and
$\tilde N=O(N d_Z/d_X)$.  The value of $\log(\tilde d)/\log(\tilde N)$ is minimized when $d_Z$ is as large as possible, i.e., when $d_Z=N$.  In this case, $\tilde d=N$ and $\tilde N=N^2/d_X$.
However, then $d_X=\Omega(N^{\mu-1})$, and so $\tilde N=O(N^{3-\mu})$.
\end{proof}
\end{corollary}

\section{Soundness}
\label{soundsec}
In this section, we consider the effect of these qubit and stabilizer splitting steps on the soundness of a quantum code, as in definition \ref{soundness}.
Because the only application of the results here that we are interested in is to codes $C$ with $w_X,w_Z,q_X,q_Z$ all $O(\log(N))$, we will only keep track of the effect up to polylogarithmic factors.  This will significantly simplify the proofs.  For such codes $C$, at every step of the construction of theorem \ref{mainth}, the codes $\tilde C$ all have
$\tilde w_X,\tilde w_Z,\tilde q_X,\tilde q_Z$ all at most polylogarithmic in $\tilde N$, so we will also assume that that holds in the bounds here.

We will also consider another concept, that we call``cosoundness".  This concept is a natural one related to soundness and so we study it here also; however, the proofs of the soundness parameters of $\tilde C$ are independent of the proofs of the cosoundness parameters of $\tilde C$ and so the reader can skip all references to cosoundness if desired.
First, let us repeat the definition of soundness in the language of chain complexes.
Consider a  CSS stabilizer code defined from a chain complex $\ldots C_{q+1} \stackrel{\partial_{q+1}}{\rightarrow} C_q \stackrel{\partial_q}{\rightarrow} C_{q-1} \ldots$, with the qubits associated with $q$-cells and the $Z$-type and $X$-type stabilizers associated with $(q+1)$-cells and $(q-1)$-cells, respectively.
Given a vector $v$, let $\wt(v)$ denote the number of nonzero entries of $v$.
Then, define soundness parameters $\epsilon_X(w),\epsilon_Z(w)$ by:
\begin{definition}
Define
\be
\epsilon_Z(w)={\rm inf}_{v\in C_q,\wt(v)=w, \partial_q v \neq 0} \Bigl( {\rm max}_{u \in C_q, \partial_q u=0} \frac{\wt(\partial v)}{\wt(v+u)}\Bigr).
\ee
Define $\epsilon_X(w)$ similarly, with $\partial_q$ replaced with $\partial_{q+1}^T$, where the superscript $T$ denotes transpose.
Let $\epsilon_Z={\rm min}_{w\neq 0} \epsilon_Z(w)$ and let $\epsilon_X={\rm min}_{w\neq 0} \epsilon_Z(w)$.
\end{definition} 

Now we define the cosoundness parameters, $\epsilon^X(w),\epsilon^Z(w)$:
\begin{definition}
\label{csounddef}
Define
\be
\epsilon^Z(w)={\rm inf}_{v\in C_{q+1},\wt(v)=w, \partial_q v \neq 0} \Bigl( {\rm max}_{u \in C_{q+1}, \partial_q u=0} \frac{\wt(\partial v)}{\wt(v+u)}\Bigr).
\ee

Define
\be
\epsilon^X(w)={\rm inf}_{v\in C_{q-1},\wt(v)=w, \partial_q v \neq 0} \Bigl( {\rm max}_{u \in C_{q-1}, \partial^T_q u=0} \frac{\wt(\partial^T v)}{\wt(v+u)}\Bigr).
\ee
Let $\epsilon^Z={\rm inf}_{w\neq 0} \epsilon^Z(w)$ and let $\epsilon^X={\rm min}^{w\neq 0} \epsilon_Z(w)$.
\end{definition} 
In terms of the generators of the code, the cosoundness $\epsilon^Z$ is related to the following question: given a product of $Z$-type stabilizer generators such that their product acts on a small number of qubits, can this product be expressed using a small number of stabilizer generators?  Indeed, if the product acts on $k$ qubits, then it can be expressed as a product of at most $\epsilon k$ generators.

We will use both definitions of soundness and cosoundness (both in terms of vectors in the chain complex and in terms of Pauli operators), depending upon what is notationally more convenient.

As an example of a code that is sound but not cosound, consider a classical repetition code using generators defined on an expander graph as follows.  Choose a graph $G$ with bounded degree and with good expansion properties.  There will be one qubit per vertex, and for each edge there is one $Z$-type generator; this generator is $Z_i Z_j$ where $i,j$ are the qubits corresponding to the vertices attached to the edge.  All $Z$-type operators commute with all the generators since there are no $X$-type generators; hence, $\epsilon_Z$ is the infimum of an empty set which is conventionally defined to be $+\infty$ (of course, it is a matter of definition how one defines the $\epsilon_Z$ in this case, we have simply given one definition).
The soundness parameter $\epsilon_X$ is lower bounded by the expansion properties of the graph.
However, if one consider a pair of vertices $i,j$ such that the shortest path between those vertices is a path $i,i_1,i_2,\ldots,i_l,j$, then the product of stabilizers
$(Z_i Z_{i_1}) (Z_{i_1} Z_{i_2}) \ldots (Z_{i_l} Z_j)=Z_i Z_j$ has weight two but cannot be expressed as a product of fewer than $l+1$ stabilizers.  Hence, the cosoundness can be inverse in the diameter of the graph and hence can be inverse logarithmic in $N$. 
We invite the reader to construct an example with constant soundness but polynomially small cosoundness or to show that it cannot be done.

Now, we prove that
\begin{lemma}
\label{stabsplitlemmaeps}
Suppose that a code $C$ was $w_X,w_Z,q_X,q_Z$ all $O({\rm polylog}(N))$. Then, the code $\tilde C$
resulting from the construction of lemma \ref{stabsplitlemma} has the property that for each soundness or cosoundness parameter $\epsilon$ (i.e., $\epsilon$ is chosen to be any of $\epsilon_X,\epsilon_Z,\epsilon^X,\epsilon^Z$) , the corresponding $\tilde \epsilon$ (i.e., $\tilde \epsilon$ is $\tilde \epsilon_X,\tilde \epsilon_Z,\tilde \epsilon^X,\tilde \epsilon^Z$, respectively) obeys
\be
\tilde \epsilon \geq \Omega(1/{\rm polylog}(N)) {\rm min}(\epsilon,1).
\ee
\begin{proof}
First we show $\tilde \epsilon_X=\epsilon_X$.  Let $O$ be a product of $X$-type operators which does not commute with at least one generator.  We can multiply $O$ by some operator $Q$ which is product of stabilizers so that the product operator $OQ$ does not act on the cut qubits.  Then, by the definition of $\epsilon_X$, there is some $X$-type operator $P$ which does not act on the cut qubits and which commutes with all generators of $C$ such that the number of generators of code $C$ which do not commute with $OQ$ is at least $\epsilon_X \wt(OPQ)$.  
Further, the number of generators of $C$ which do not commute with $OQ$ is the same as the number of generators of $\tilde C$ which do not commute with $O$.  Hence,
the number of generators of code $\tilde C$ which do not commute with $O$ is at least $\epsilon_X \wt(OPQ)$.

Next, consider $\tilde \epsilon_Z$.  Let $O$ be a $Z$-type operator which does not commute with at least
$G$ generators of $\tilde C$.
Let $T_s$ denote the set $T$ constructed in some splitting step, where $s$ labels the particular splitting step.
Suppose for some $s$, $O$ does not commute with the product of operators in $T_s$, so that $O$ does not commute with the generator that was split in that step.  Then, $O$ does not commute with at least one operator in $T_s$.
Let $O=A B$ where $A$ acts on the non-cut qubits and $B$ acts on the cut qubits.
By definition of $\epsilon_Z$, there is some product $P$ of $Z$-type stabilizers in $C$ such that
the number of generators of $C$ that do not commute with $A$ is at least equal to $\epsilon_Z \wt(PA)$.
Let $\tilde P$ be the product of the matching $Z$-type stabilizers in $\tilde C$.
Let $\wt_{nc}(\ldots)$ denote the weight of an operator on the non-cut qubits; i.e., it is the number of non-cut qubits
that it acts on.  Thus, the number of generators of $C$ that do not commute with $A$ is at least
equal to $\epsilon_Z \wt_{nc}(\tilde PA)$, and so
\be
\label{e1}
G \geq \epsilon_Z \wt_{nc}(\tilde P O).
\ee
 Label cut qubits constructed in a given splitting step by a pair $(s,a)$ where $s$ labels the step and $a$ labels the cut qubit $(a)$.
Now, if for some $s$, $\tilde PO$ acts on some qubit $(s,a)$ for at least one $a$ and $O$ commutes with all of the
generators in $T_s$, then $\tilde P O$ acts on at least one of the qubits $q_1, \ldots, q_w$ in the generator $X_{q_1} \ldots X_{q_w}$ that was splitting in that step.
Let $\wt_c(\ldots)=\wt(\ldots)-\wt_{nc}(\ldots)$, i.e., $\wt_c(\ldots)$ is the number of cut qubits that an operator acts on.
The ratio $\wt_c(\tilde P O)/w_X$ is a lower bound on the number of $s$ such that $\tilde PO$ acts on at least one qubit $(s,a)$ and so there are at least 
$\wt_c(\tilde P O)/w_X - G$ choices of $s$ such that $\tilde PO$ acts on at least one qubit $(s,a)$ and commutes with all generators in $T_s$; for these $s$, $\tilde PO$ acts on at least one noncut qubit which is acted on by the stabilizer split in that step.
So,
\be
\label{e2}
\wt_{nc}(\tilde P O) \geq \frac{1}{q_X}(\wt_c(\tilde P O)/w_X - G).
\ee
Hence, by Eqs.~(\ref{e1},\ref{e2}),
after some algebra we find that $G \geq \epsilon_Z (1+w_X(\epsilon_Z+q_X))^{-1} \wt(\tilde PO)$.

Now consider $\tilde \epsilon^X$.  Let $T_s$ denote the set $T$ constructed in some splitting step, where $s$ labels the particular splitting step.
Consider some operator $O$ which is a product of $X$-type generators of $\tilde C$.
Let us write $O$ as a product of operators $F U_{s_1} U_{s_2}... U_{s_j}$, where $F$ does not act on the cut qubits (i.e., $F$ is a product of the generators of $C$) and each operator $U_{s_a}$ is a nonempty product of
some subset of the set $T_{s_a}$ and $U_{s_a}$ is not the product of all operators in $T_a$ (if it were the product of all such operators, the $U_{s_a}$ would not act on the cut qubits).
By assumption, $F$ can be written as a product of at most $\epsilon^X \wt(F)$ generators of $C$ and hence
it can be written as at most $\epsilon^X \wt(F) w_X$ generators of $\tilde C$.
Each operator $U_{s_a}$ can be written as a product of at most $w_X$ generators of $\tilde C$ and
each $U_{s_a}$ acts on at least $1$ cut qubit and at most $w_X$ non-cut qubits.
Thus, the number of non-cut qubits that $O$ acts on is at least $\wt(F)-w_X j$ and the number of cut qubits that
$O$ acts on is at least $j$, so
$\wt(O)\geq {\rm max}(\wt(F)-w_X j,j)$.
So, $\wt(F)\leq \wt(O)(w_X+1)$.  So, $O$ is a product of at most $\epsilon^X w_X(w_X+1) \wt(O)$ generators of $\tilde C$.
Also, the product $U_{s_1} U_{s_2}... U_{s_j}$ is a product of at most $w_X j \leq w_X \wt(O)$ generators of $\tilde C$.  Hence, $\tilde \epsilon^X \leq \epsilon^X w_X (w_X+1) + w_X$.

Finally consider $\tilde \epsilon^Z$.  Let $\tilde O$ be any product of $Z$-type generators of $\tilde C$.
By definition of $\epsilon^Z$, the matching product  $O$ of generators of $C$ can be be written
as a product $P$ of at most $\epsilon^Z \wt_{nc}(O)\leq \epsilon^Z \wt(O)$ generators of $C$.  Let $\tilde P$ be the product of generators
matching $P$.  We claim that $\tilde O=\tilde P$; indeed, this follows since $O=P$ and so $OP$ equals identity and so $\tilde O \tilde P$ does not act on the non-cut qubits and so $\tilde O \tilde P$ equals identity since it commutes with
all $X$-type generators.  Note that it is not possible for $\wt_{nc}(O)=0$ and $\wt(O) \neq 0$ as $O$ commutes with
all $X$-type generators.
\end{proof}
\end{lemma}

\begin{lemma}
\label{prodCeps}
Suppose that a code $C$ was $w_X,w_Z,q_X,q_Z$ all $O({\rm polylog}(N))$. 
Use the values of $l,w$ from Eq.~\ref{lchoicehere},\ref{wchoicehere}).
Then, the code
resulting from the construction of lemma \ref{prodC} has  the property that for each soundness or cosoundness 
parameter $\epsilon$ (i.e., $\epsilon$ is chosen to be any of $\epsilon_X,\epsilon_Z,\epsilon^X,\epsilon^Z$) , the corresponding $\tilde \epsilon$ (i.e., $\tilde \epsilon$ is $\tilde \epsilon_X,\tilde \epsilon_Z,\tilde \epsilon^X,\tilde \epsilon^Z$, respectively) obeys
\be
\tilde \epsilon \geq \Omega(1/{\rm polylog}(N)) {\rm min}(\epsilon,1).
\ee
\begin{proof}
First consider $\tilde \epsilon_Z$.  
Let $O$ be a $Z$-type operator.  Suppose $O$ does not commute with $G$ generators.
We claim that without loss of generality we can assume that $O$ acts only on qubits labelled $(q,1)$ and $[s_X,k]$ and not on qubits $(q,k)$ for $k>1$.    
This holds because we can multiply $O$ by $Z$-type generators
of the form $Z_{(q,k)} Z_{(q,k+1)} \prod_{s_X \ni q} Z_{[s_X,k]}$ to achieve this.
Let $\wt_1(\ldots)$ denote the weight of an operator on qubits of the form $(q,1)$.
Let $\wt_{\perp}(\ldots)=\wt(\ldots)-\wt_1(\ldots)$.
Now, since $O$ does not acts on qubits $(q,k)$ for $k>1$, if $O$ does act on at least one qubit $[s_X,k]$ for some $s_X$, it must fail to commute with at least one of the generators 
$$X_{(q_1,k)} X_{(q_2,k)} \ldots X_{(q_w,k)} X_{[s_X,k-1]} X_{[s_X,k]}$$ for $k=2,\ldots,l-1$ or
$$X_{(q_1,k)} X_{(q_2,k)} \ldots X_{(q_w,k)} X_{[s_X,k-1]}$$ for $k=l$.
Hence,
\be
\label{ec1}
G \geq \wt_\perp(O)/l.
\ee
By definition of $\epsilon_Z$, the number of $X$-type operators of the form
$X_{(q_1,k)} X_{(q_2,k)} \ldots X_{(q_w,k)}$, where $X_{q_1} \ldots X_{q_w}$ is a generator of $C$, that
do not commute with $O$ is at least equal to $\epsilon_Z \wt_1(O)$ and so 
\be
\label{ec2}
G \geq \epsilon_Z \wt_1(O)-\wt_\perp(O).
\ee
So, by Eq.~(\ref{ec1},\ref{ec2}) and since $\wt(O) \geq {\rm max}(\wt_\perp(O),\wt_1(O))$, we have $G\geq 
\wt(O)/(l+1)$.

Now consider $\tilde \epsilon_X$.
Let $u \in C_2 \otimes E_0 \oplus C_1 \otimes E_1$.  Let $u=v+y$ with $v \in C_2 \otimes E_0$ and $y \in C_1 \otimes E_1$.
Without loss of generality, we can assume that $y=0$; if not, let $y=\sum_i y_i \otimes e_i$, where $e_i$ for $i=1,l-1$ are standard basis vectors for $E_1$ and 
add $\partial_2^T (y_1 \otimes w_1 + (y_1+y_2) \otimes w_2 + (y_1+y_2+y_3) \otimes w_3 + \ldots)$ to $y$, where $w_i$ are standard basis vectors for $E_0$.
Let $v=\sum_i v_i \otimes w_i$.
By definition of soundness, without loss of generality we can assume that $\wt(\partial_3^T v_1) \geq \epsilon_X \wt(v_1)$; to see this, if this is not true, then
there exists $x \in C_2$ such that $\partial_3^T x=0$ and $\wt(\partial_3^T v_1) \geq \epsilon_X \wt(v_1+x)$ and we can
add $x \otimes \sum_i w_i$ to $u$.
So,
\be
\label{es1}
\wt(\partial_3^T u) \geq \epsilon_X \wt(v_1).
\ee
However, considering the weight of $u$ in $C_2 \otimes E_1$ (i.e., the number of nonzero entries of $u$ projected into that subspace), we have that
$\wt(\partial_3^T u)\geq {\rm max}_k \wt(v_k-v_1) \geq {\rm mak}_k \wt(v_k) -\wt(v_1))$ and so
\be
\label{es2}
\wt(\partial_3^T u) \geq \wt(v)-l \wt(v_1).
\ee
So, by Eqs.~(\ref{es1},\ref{es2}), $\wt(\partial_3^T u) \geq \epsilon_X \wt(v)/(\epsilon_X+l)$.

Now consider $\tilde \epsilon^X$.  Let $u \in C_1 \otimes E_0$.
Let $u=\sum_i u_i \otimes w_i$, with $w_i$ standard basis vectors for $E_0$.
Without loss of generality, we can assume that $\wt(\partial_2^T u_1) \geq \epsilon^X \wt(u_1)$; if not, there exists $x\in C_1$ such that $\partial_2^T x=0$ and
$\wt(\partial_2^T x) \geq \epsilon^X \wt(u_1+x)$ and we can add $u\otimes \sum_k w_k$ to $u$.
So,
\be
\label{et1}
\wt(\partial_2^T u) \geq \epsilon^X \wt(u_1).
\ee
Also, by considering the weight of $\partial_2^T u$ in $C_1 \otimes E_1$, we have
that $\wt(\partial_2^T u) \geq {\rm max}_k \wt(v_k)-\wt(v_1)$ and so
\be
\label{et2}
\wt(\partial_2^T u) \geq \wt(u)-l \wt(u_1).
\ee
So, by Eqs.~(\ref{et1},\ref{et2}), $\wt(\partial_2^T u) \geq \epsilon^X \wt(u)/(\epsilon^X+l)$.

Now consider $\tilde \epsilon^Z$.  Consider a code that we call $\hat C$.  This will be a code constructed
following the construction of lemma \ref{prodC} in which we choose all of $Z$-type generators
which are of the form $Z_{(q_1,k)} \ldots Z_{(q_w,k)}$ to have $k=1$.  We will lower bound the corresponding
cosoundness parameter for code $\hat C$; we denote this parameter $\hat \epsilon^Z$.  This will imply a lower bound on $\tilde \epsilon^Z$ since $\tilde \epsilon^Z \geq \hat \epsilon^Z/l$ because any generator in $\hat C$ can be written as a product of at most
$l$ generators in $\tilde C$.  Let $u\in C_3 \otimes E_0 \oplus C_2 \otimes E_1$.  Let $u=v \otimes w_1+w$ with $v   \in C_3$ and $w_1$ being one of the standard basis vectors for $E_0$, and with $w \in C_2 \otimes E_1$.
By definition of
$\epsilon^Z$, there is some vector $x\in C_3$ such that $\partial_3 x=0$ and $\wt(\partial_3 v) \geq \epsilon^Z \wt(v+x)$.  Adding $x \otimes w_1$ to $u$, we can assume without loss of generality that
$\wt(\partial_3 v) \geq \epsilon^Z \wt(v)$.
Let $\wt_{20}(\ldots)$ denote the weight of a vector projected into $C_2 \otimes E_0$.
Note that
\be
\label{ee1}
\wt_{20}(\partial_3 (v \otimes w_1)+w) \geq \wt_{20}(\partial_3(v \otimes w_1)).
\ee
(To see this, for every basis vector $q$ of $C_2$, the parity of the number of nonzero entries of
$\partial_3 (v \otimes w_1)+w$ on qubits of the form $(q,k)$ is odd if $\partial_3 (v)$ is equal to $1$ in its $q$-th entry; that is, adding $w$ does not change this parity.)
Also, (this next equation follows because for any vector $\in E_1$, the weight of $\partial'$ of that vector is at least $1/l$ times the weight of that vector)
\be
\label{ee2}
\wt_{20}(\partial_3 w) \geq \wt(w)/l.
\ee
Also,
\be
\label{ee3}
\wt_{20}(\partial_3 (v\otimes w_1+w)) \geq \wt_{20}(\partial_3 w)-wt_{20}(\partial_3(v \otimes w_1))\geq
\wt(w)/l-\wt_{20}(\partial_3(v \otimes w_1))
\ee
So, by Eqs.~(\ref{ee1},\ref{ee3}),
\be
\label{ee4})
\wt_{20}(\partial_3 (v \otimes w_1)+w)  \geq \wt(w)/(2l).
\ee
Using Eq.~(\ref{ee1}) again, we have $\wt_{20}(\partial_3 (v \otimes w_1)+w) \geq \epsilon^Z \wt(v)$.
So,
$\wt_{20}(\partial_3 (v \otimes w_1)+w) \geq {\rm max}(\epsilon^Z \wt(v),\wt(w)/(2l))$.
Since $\wt(v)+\wt(w)=\wt(u)$,
after some algebra we find
$\wt_{20}(\partial_3 (v \otimes w_1) \geq \epsilon^Z (1+2l\epsilon^Z)^{-1} \wt(u)$.
\end{proof}
\end{lemma}

\section{Alternative Qubit Splitting}
\label{altsplit}
In this section we give an alternative way to reduce $q_X$.  This method does not work as well as the method of lemma \ref{prodC}, because it can lead to a large $\tilde w_Z$.  Geometrically, this step can be interpreted as follows: a qubit was a $2$-cell with the topology of a multiply punctured sphere.  Imagine cutting this two cell into two halves, northern and southern hemispheres, so that half the punctures are in each hemisphere, with no punctures crossing the equator.  This cutting adds an additional $1$-cell (an additional $X$-type stabilizer) at the equator, with both hemispheres attached to this cell.  Repeat this cutting process several times until only $3$ punctures are in each sphere.  We now explain the step in terms of its action on stabilizers.

We define an ``alternative $X$-type qubit splitting step" is defined by choosing a qubit $q$.  Let $w$ denote the number of $X$-type generators acting on qubit $q$.  We then define a new code with $N+w-1$ qubits labelled $1,\ldots,q-1,q+1,\ldots,N$ and $q(1),q(2),\ldots,q(w)$.  This code has $n_X+w-1$ $X$-type generators as follows.  $w-1$ of these generators are
$X_{q(a)} X_{q(a+1)}$ for $a=1,\ldots,w-1$.
The other $n_X$ $X$-type generators are derived from the $X$-type generators of the old code.  For each $Z$-type generator $R$ of the old code, we define a generator $R'$ of the
 new code.  If $R$ does not act on $q$, then $R'=R$.  If $R$ does act on $q$, then we replace $q$ with one of the $q(a)$ so that each $q(a)$ has one of these generators acting on it.
The new code has $n_Z$ $Z$-type generators.  For each $Z$-type generator $R$ of the old code, we define a generator $R'$ of the new code.  If $R$ does not act on $q$ then $R'=R$.  If $R$ does act on $q$, then we define $R'$ by replacing $Z_q$ with $Z_{q(1)} \ldots Z_{q(w)}$.
This step has the effect of reducing the number of $X$-type generators acting on a qubit, so that the number acting on any $q(a)$ is at most $3$.  However, the weight of at most
$w q_Z$
$X$-type generators is increased by $1$.

Given a code $C$, we then define a new code $\tilde C$ by applying this alternative splitting step to all qubits in $C$.
The resulting code has the following properties:
\begin{lemma}
\begin{itemize}
\item[1.] $\tilde q_X\leq 3$.

\item[2.] $\tilde q_Z=q_Z$.

\item[3.] $\tilde w_X=w_X$.

\item[4.] $\tilde w_Z\leq w_Z q_X$.

\item[5.] $\tilde K=K$.
\end{itemize}
\begin{proof}
Immediate.
\end{proof}
\end{lemma}

\section{Discussion}
We have given a method to construct strongly LDPC quantum code families from other quantum code families.  The increase in number of physical qubits and the effect on distance depends upon the parameters of the original code, in a way described above.

The construction is based on several different steps to reduce the parameters.
It is possible that there are more optimal ways to apply these steps, by changing the order.
For example, perhaps one might reduce the weight of some, but not all stabilizers, or only partially reduce the weight of stabilizers, before applying the qubit splitting step, and then follow with further reduction of stabilizer weight.  Such alternative procedures might also allow a more balanced treatment of $X$ and $Z$, so that if $C$ has $d_X=d_Z$ then $C^{new}$ has $d_X=d_Z$ without needing to balance the code at the end.

It may also be possible to perform the stabilizer splitting step in a different manner, roughly as follows.  The $Z$-type stabilizer splitting step has the interpretation (considering for this paragraph a different geometric interpretation where the $Z$-type stabilizers are $2$-cells, not $3$-cells) that we start with a $w$-side polygon, where $w$ is the weight of the stabilizer, and we divide it into $O(w)$ triangle.  This division into triangles involves adding extra $1$-cells (extra qubits) but not extra $0$-cells (extra $X$-type stabilizers).  However, one might instead add additional $0$-cells inside the polygon when splitting it.  This would lead perhaps to an increase in $\tilde N$, which is undesirable, but might lead to better distance properties.

{\it Acknowledgments---} I thank J. Haah for useful discussions.


\begin{thebibliography}{99}
\bibitem{review} Eq.~\ref{Khom} is in used in many of the references above \cite{csshomology1,csshomology2,csshomology3}.  See also Ref.~\onlinecite{bh} for a review of a ``dictionary" between the language of logical operators and homology.
\end{thebibliography}
\end{document}